\documentclass[reprint,prd,aps,preprintnumbers,tightenlines,nofootinbib,superscriptaddress
]{revtex4-2}

\usepackage{graphicx} 
\usepackage[usenames,dvipsnames]{xcolor}
\usepackage{xspace} 
\usepackage[utf8]{inputenc}
\usepackage[normalem]{ulem}
\usepackage{graphicx}
\usepackage{mathrsfs,amsmath,amsfonts,amsthm,bm}

\xdefinecolor{mylinkcolor}{rgb}{0,0,0.5}
\usepackage[
	bookmarksnumbered, bookmarksopen, bookmarksopenlevel=2,
	breaklinks=true,
	colorlinks=true, filecolor=mylinkcolor, citecolor=mylinkcolor,
	linkcolor=mylinkcolor, urlcolor=mylinkcolor, menucolor=mylinkcolor,
]{hyperref}
\usepackage{orcidlink}

\allowdisplaybreaks

\DeclareMathOperator{\Order}{\mathcal{O}}
\newcommand{\mr}{\mathrm}

\newcommand{\PSstr}{$\text{PS}^*~$}

\newcommand{\AEI}{\affiliation{Max Planck Institute for Gravitational Physics (Albert Einstein Institute), Am M\"uhlenberg 1, Potsdam 14476, Germany}}
\newcommand{\Maryland}{\affiliation{Department of Physics, University of Maryland, College Park, MD 20742, USA}}

\begin{document}

\title{Energetics and scattering of gravitational two-body systems at fourth post-Minkowskian order}

\author{Mohammed Khalil\,\orcidlink{0000-0002-6398-4428}}\email{mohammed.khalil@aei.mpg.de}\AEI\Maryland
\author{Alessandra Buonanno\,\orcidlink{0000-0002-5433-1409}}\email{alessandra.buonanno@aei.mpg.de}\AEI\Maryland
\author{Jan Steinhoff\,\orcidlink{0000-0002-1614-0214}}\email{jan.steinhoff@aei.mpg.de}\AEI
\author{Justin Vines\,\orcidlink{0000-0001-6471-5409}}\email{justin.vines@aei.mpg.de}\AEI

\begin{abstract}
Upcoming observational runs of the LIGO-Virgo-KAGRA collaboration, and future gravitational-wave (GW) detectors on the ground and in space, require waveform models more accurate than currently available. 
High-precision waveform models can be developed by improving the analytical description of compact binary dynamics and completing it with numerical-relativity (NR) information. 
Here, we assess the accuracy of the recent results for the fourth post-Minkowskian (4PM) conservative dynamics through comparisons with NR simulations for the circular-orbit binding energy and for the scattering angle.
We obtain that the 4PM dynamics gives better agreement with NR than the 3PM dynamics, and that while the 4PM approximation gives comparable results to the third post-Newtonian (PN) approximation for bound orbits, it performs better for scattering encounters. 
Furthermore, we incorporate the 4PM results in effective-one-body (EOB) Hamiltonians, which improves the disagreement with NR over the 4PM-expanded Hamiltonian from $\sim 40\%$ to $\sim 10\%$, or $\sim 3\%$ depending on the EOB gauge, for an equal-mass binary, two GW cycles before merger. 
Finally, we derive a 4PN-EOB Hamiltonian for hyperbolic orbits, and compare its predictions for the scattering angle to NR, and to the scattering angle of a 4PN-EOB Hamiltonian computed for elliptic orbits.
\end{abstract}

\maketitle

\section{Introduction}

Gravitational-wave (GW) observations~\cite{LIGOScientific:2016aoc,LIGOScientific:2021djp,LIGOScientific:2020ibl,
LIGOScientific:2018mvr} have improved our understanding of compact binaries, composed of 
black holes and/or neutron stars, their properties, and their astrophysical formation
channels~\cite{LIGOScientific:2018jsj,LIGOScientific:2020kqk,LIGOScientific:2021psn}.
Searching for GW signals and estimating their parameters require accurate 
waveform models or templates.  Since numerical-relativity (NR) simulations are
computationally expensive, analytical approximation methods become
essential for producing such waveforms.

\begin{figure*}
\centering
\includegraphics[width=0.9\linewidth]{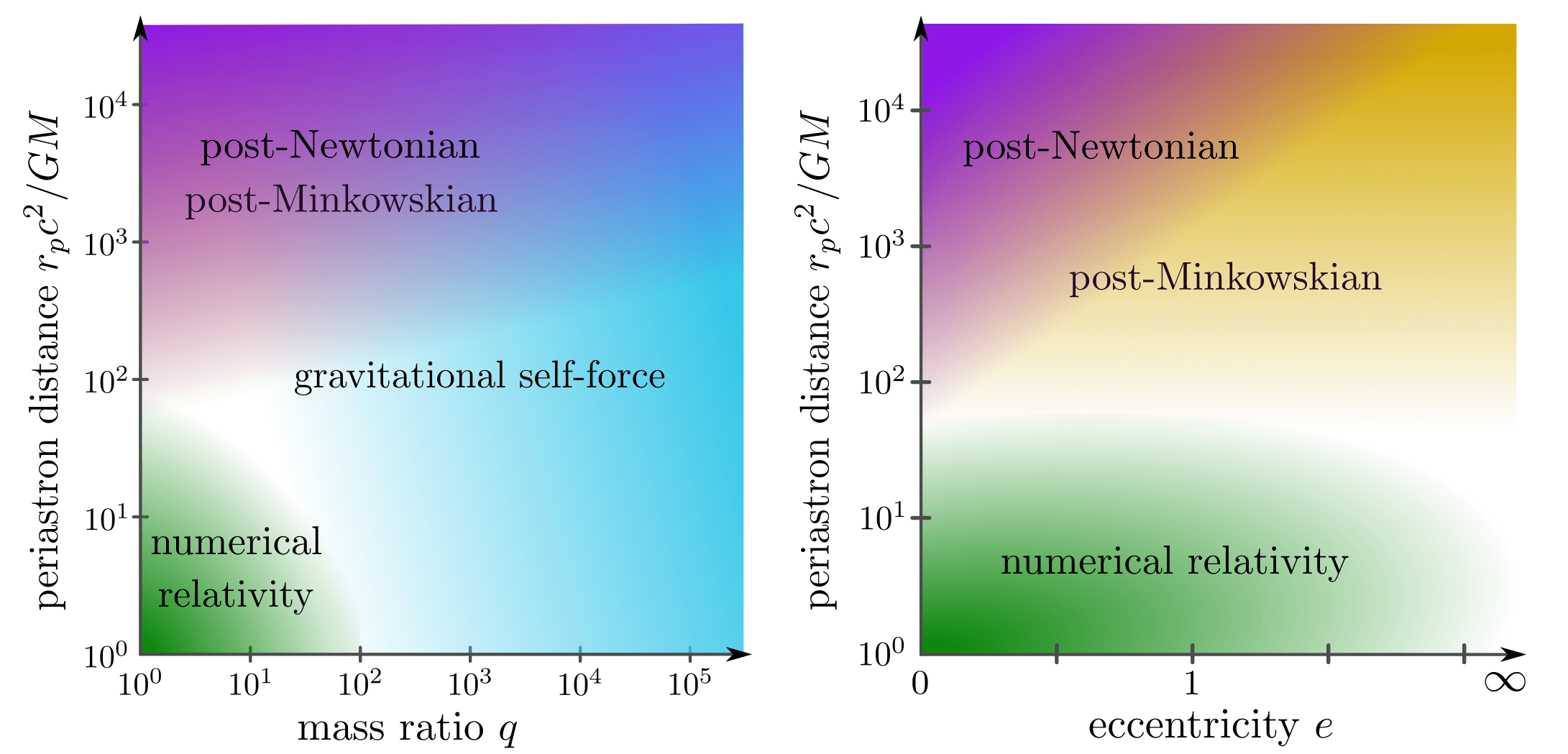}
\caption{The left panel shows the region of applicability of NR and the PN, PM, and GSF approximations for small eccentricity $e \sim 0$, in which case the PN and PM approximations overlap.
The right panel shows the range in eccentricity for which each approximation is applicable for comparable masses $q \sim 1$. The PM approximation is more accurate than the PN approximation for scattering encounters at large velocities, or equivalently large eccentricities at fixed periastron distance.}
\label{fig:PMPN} 
\end{figure*}

The post-Newtonian (PN) approximation is valid for slow motion,
$v^2/c^2 \ll 1$, and weak gravitational potential, $GM/rc^2 \ll 1$,
making it most applicable for binaries in bound orbits where $v^2/c^2
\sim GM/rc^2$, which are the main sources observed by ground-based GW
detectors, such as LIGO, Virgo and KAGRA~\cite{LIGOScientific:2014pky,VIRGO:2014yos,KAGRA:2020agh}; for reviews, see
Refs.~\cite{Futamase:2007zz,Blanchet:2013haa,Schafer:2018kuf,Levi:2015msa,Porto:2016pyg,Levi:2018nxp}.

Similarly, the post-Minkowskian (PM) approximation is a weak-field
expansion, but places no restriction on the magnitude of velocities.
Next to entirely classical approaches to the PM
approximation~\cite{Westpfahl:1979gu,Westpfahl:1980mk,Bel:1981be,schafer1986adm,Ledvinka:2008tk,Damour:2016gwp,Damour:2017zjx,Damour:2019lcq,Blanchet:2018yvb},
recent progress has been pioneered by methods starting out from
(quantum) scattering amplitudes~\cite{Arkani-Hamed:2017jhn,Bjerrum-Bohr:2018xdl,Kosower:2018adc,Cheung:2018wkq,Bautista:2019tdr,Bern:2019nnu,Bjerrum-Bohr:2019kec,Cristofoli:2020uzm,Bern:2021dqo,Bern:2021yeh,Travaglini:2022uwo,Bjerrum-Bohr:2022blt,Kosower:2022yvp,Buonanno:2022pgc}.
In addition, manifestly classical methods that make use of
quantum-field-theory techniques show great promise for advancing the
PM approximation, namely effective field
theory~\cite{Foffa:2013gja,Kalin:2019rwq,Kalin:2020fhe,Kalin:2020mvi,Dlapa:2021npj,Dlapa:2021vgp}
and worldline quantum field
theory~\cite{Mogull:2020sak,Jakobsen:2021smu} approaches.  The PM
approximation has also been applied to spin~\cite{Bini:2017xzy,Bini:2018ywr,Vines:2017hyw,Vines:2018gqi,Guevara:2018wpp,Guevara:2019fsj,Kalin:2019inp,Bern:2020buy,Chung:2020rrz,Maybee:2019jus,Chung:2019duq,Bautista:2021wfy,Kosmopoulos:2021zoq,Liu:2021zxr,Chen:2021qkk,Jakobsen:2021lvp,Jakobsen:2021zvh,Jakobsen:2022fcj,Bern:2022kto,Aoude:2022trd}, tidal~\cite{Bini:2020flp,Cheung:2020sdj,Kalin:2020lmz,Bern:2020uwk,Cheung:2020gbf,Haddad:2020que,Aoude:2020ygw,Mougiakakos:2022sic},
and radiative effects~\cite{Damour:2020tta,DiVecchia:2020ymx,DiVecchia:2021bdo,DiVecchia:2021ndb,Herrmann:2021tct,Bjerrum-Bohr:2021din,Bini:2021gat,Bini:2021qvf,Brandhuber:2021eyq,Damgaard:2021ipf,Herrmann:2021lqe,Riva:2021vnj,Saketh:2021sri,Cho:2021arx,Manohar:2022dea}.

The PM expansion encompasses the PN expansion, such that the $(n+1)$PM order includes all the information up to the $n$PN order, making it potentially more accurate. 
Binaries in bound orbits can reach velocities on the order of $0.4$ or larger when spiraling over the last orbits before merger. This means that the 
relativistic corrections become more and more important in the last stages of the inspiral 
and plunge. Thus, we might expect that at some high PM order, the PM expansion may start to become more accurate than the PN 
expansion. Furthermore, scattering encounters on 
hyperbolic trajectories can reach high velocities, for which the PM approximation becomes more relevant. 
The gravitational self-force (GSF) approximation~\cite{Mino:1996nk,Quinn:1996am,Barack:2001gx,Barack:2002mh,Gralla:2008fg,Detweiler:2008ft,Keidl:2010pm,vandeMeent:2017bcc,Pound:2012nt,Pound:2019lzj,Gralla:2021qaf,Barack:2018yvs,Pound:2021qin} expands the Einstein equations 
in power of the binary's mass ratio $m_2/m_1 \ll 1$. It is thus valid for any 
velocity and is not restricted to the weak fields.
Figure~\ref{fig:PMPN} illustrates the regions of parameter space in
which the PN, PM and GSF approximations are (roughly) applicable 
when generic orbits are considered.

Detecting GW bursts from hyperbolic encounters would have important
implications on our understanding of dense stellar environments and
the merger rate of compact objects formed 
through this astrophysical channel.  Currently, there does not seem to be a
consensus in the literature~\cite{Kocsis:2006hq,Ivanova:2007bu,OLeary:2008myb,Capozziello:2008ra,Tsang:2013mca,Kremer:2019iul,Mukherjee:2020hnm}
on the event rate for scattering encounters, due to the high
uncertainty in the astrophysical models; the expected event rates
vary between 0.001 to 0.39 per year for upcoming LIGO-Virgo-KAGRA 
runs, depending on the model~\cite{Mukherjee:2020hnm}, with 
higher rates expected for future detectors. Gravitational waves from hyperbolic encounters would be expected at higher
rates in the LIGO-Virgo-KAGRA frequency band if a large population of primordial
black holes exists in dense stellar clusters, as was proposed in
Refs.~\cite{Clesse:2015wea,Garcia-Bellido:2017qal,Garcia-Bellido:2017knh,Morras:2021atg}
based on some inflationary models. Other GW sources that could reach highly-relativistic velocities are binaries in galactic nuclei, where dynamical capture and three-body interactions can drive binaries to high eccentricities~\cite{OLeary:2008myb,Gondan:2020svr,Gultekin:2005fd,Samsing:2013kua,McKernan:2012rf,VanLandingham:2016ccd,Stone:2016wzz,Leigh:2017wff,Repetto:2017gry,Hamers:2018hxv,Rodriguez:2018pss,Britt:2021dtg}.

Most studies of hyperbolic and parabolic encounters use the PN approximation, for both the dynamics~\cite{damour1985general,Cho:2018upo,Bini:2012ji,Bini:2021jmj} and the GW energy spectrum~\cite{Capozziello:2008ra,Berry:2010gt,DeVittori:2012da,DeVittori:2014psa,Grobner:2020fnb}.
The effective-one-body (EOB) formalism~\cite{Buonanno:1998gg,Buonanno:2000ef} has also been applied to scattering, as in Refs.~\cite{Damour:2014afa,Nagar:2020xsk,Nagar:2021gss,Nagar:2021xnh,Gamba:2021gap,Ramos-Buades:2021adz}.
EOB waveform models improve the inspiral-merger-ringdown waveforms by combining test-body, PN, black-hole perturbation, and NR results.
PM results have been incorporated in EOB Hamiltonians in Refs.~\cite{Damour:2016gwp,Damour:2017zjx,Antonelli:2019ytb,Damgaard:2021rnk}.

In this paper, we study the results of 
  Refs.~\cite{Bern:2021dqo,Bern:2021yeh,Dlapa:2021npj,Dlapa:2021vgp}
  for the 4PM conservative dynamics, both the ``vanilla'' 4PM Hamiltonians and the 
EOB Hamiltonians constructed with 4PM conservative information.  Generally, the waveform models are evaluated on the 
two-body dynamics, which is derived from the Hamiltonian and the radiation-reaction (RR) force. 
  The Hamiltonian describes the
  conservative dynamics, while the RR force accounts for the
  energy and angular momentum losses due to GWs, and it is included in
  the right-hand side of the Hamilton equations. Since the RR force has not 
been computed in PM theory at sufficiently high PM order, in this
  paper, we restrict our study to the conservative sector, and
  therefore focus on the Hamiltonian. As shown in
  Refs.~\cite{Barausse:2011dq,Damour:2011fu,Nagar:2015xqa}, a good
  diagnostic to assess the accuracy of models for the
  conservative dynamics is to compare the binding energy from the
  Hamiltonians and NR. The latter contains both conservative and dissipative
  effects. When augmenting the models' dynamics with RR effects, the
  binding energy can change (e.g., see Fig. 6 in
  Ref.~\cite{Antonelli:2019ytb}).  However, it is still very informative to
  perform studies that compare to NR only the conservative
  dynamics at various orders of the perturbative expansion, and also 
develop different flavors of Hamiltonians, and explore their closeness 
to NR. Eventually, when RR effects in PM theory become available, one will 
be able to identify the most suitable model for the full dynamics that best represent the 
NR results. Quite interestingly, in the case of the
  scattering angle, for which we can add the RR effects to the models'
  predictions, we show that the radiative
  contribution is much smaller than the conservative one (see Fig.~\ref{fig:chiPM} below).

The paper is structured as follows. 
In Sec.~\ref{sec:PMHam} and Sec.~\ref{sec:EOBHam}, we summarize the 4PM results and incorporate them in EOB Hamiltonians. 
Then, in Sec.~\ref{sec:energy}, we compare the circular-orbit binding energies with NR, while in Sec.~\ref{sec:scattering}, we compare the scattering angles.
We conclude in Sec.~\ref{sec:conc}, and write in Appendices~\ref{app:Hiso} and \ref{app:eobcoefs} the expressions for the PN and EOB Hamiltonians. We also provide those expressions as a \textsc{Mathematica} file in the Supplemental Material~\cite{ancmaterial}.

\subsection*{Notation}
We adopt units in which the speed of light $c = 1$.

For a binary with masses $m_1$ and $m_2$, with $m_1 \geq m_2$, we define the following quantities:
\begin{gather}
M\equiv m_1 + m_2, \quad \mu \equiv \frac{m_1m_2}{M}, \quad \nu \equiv \frac{\mu}{M}, \quad
q \equiv \frac{m_1}{m_2}.
\end{gather}

From the total energy $E$, we define the binding energy:
\begin{equation}
\bar{E} \equiv \frac{E - M}{\mu},
\end{equation}
and the effective energy $E_\text{eff}$ through the EOB energy map~\cite{Buonanno:1998gg}
\begin{equation}
E = M \sqrt{1 + 2\nu \left(\frac{E_\text{eff}}{\mu} - 1\right)}\,.
\end{equation}
We also introduce the following quantities:
\begin{align}
\gamma &\equiv \frac{E_\text{eff}}{\mu}= \frac{E^2 - m_1^2 - m_2^2}{2 m_1 m_2}, \nonumber\\
\Gamma &\equiv \frac{E}{M} = \sqrt{1 + 2 \nu (\gamma -1)}\,,
\end{align}
where $\gamma$ is related to the asymptotic relative velocity $v$ by
\begin{equation}
v \equiv \frac{\sqrt{\gamma^2 - 1}}{\gamma},  \quad \text{or} \quad \gamma = \frac{1}{\sqrt{1-v^2}}.
\end{equation} 
When dealing with PN expansions, it is convenient to define the dimensionless energy variable\footnote{
Note that $\varepsilon$ used here is denoted $p_\infty^2$ in Refs.~\cite{Bini:2020wpo,Bern:2021yeh}.}
\begin{equation}
\varepsilon \equiv \gamma^2 - 1 = \gamma^2 v^2.
\end{equation}

The orbital angular momentum is denoted $L$, and is related to the relative position $R$, radial momentum $P_R$, and total linear momentum $P$ via 
\begin{equation}
P^2 = P_R^2 + \frac{L^2}{R^2}.
\end{equation}

We use $H_\text{hyp}$ to denote a Hamiltonian with PM or PN information that is valid for unbound/hyperbolic motion, and use $H_\text{ell}$ for a Hamiltonian valid for bound/elliptic orbits. A Hamiltonian written without a `hyp' or `ell' subscript is valid for generic motion.

We often use the following dimensionless rescaled quantities:
\begin{gather}
r\equiv\frac{R}{GM}, \quad  u \equiv \frac{1}{r},  \quad p \equiv \frac{P}{\mu},  \quad p_r \equiv \frac{P_R}{\mu},  \nonumber\\
l \equiv\frac{L}{GM\mu}, \quad \hat{H} \equiv \frac{H}{\mu}.
\end{gather}

\section{PM-expanded scattering angle and Hamiltonian}
\label{sec:PMHam}

The 4PM conservative dynamics (including tail effects) has been
derived recently in
Refs.~\cite{Bern:2021dqo,Bern:2021yeh,Dlapa:2021npj,Dlapa:2021vgp} for
\emph{hyperbolic orbits} in a large-eccentricity expansion. We note
that this 4PM result agrees with the 6PN result of
Refs.~\cite{Bini:2021gat,Bini:2020hmy,Bini:2020nsb}, and exhibits a
simple mass dependence, which is expected due to Lorentz invariance
and dimensional analysis, as argued in Ref.~\cite{Damour:2019lcq}. The
result of
Refs.~\cite{Bern:2021dqo,Bern:2021yeh,Dlapa:2021npj,Dlapa:2021vgp}
also agrees with the 5PN result of
Refs.~\cite{Blumlein:2021txe,Blumlein:2020pyo,Almeida:2021xwn}, except
for a single term that does not have the expected mass dependence, and
is proportional to $\nu^2$.~\footnote{ The difference in the 5PN
  conservative scattering angle between
  Refs.~\cite{Bini:2020wpo,Bern:2021yeh} and \cite{Blumlein:2021txe},
  which is given by Eq.~(69) of the latter, is proportional to $\nu^2
  v^6 / L^4$. In all configurations considered in this paper, the
  velocities reached by a binary system are typically $\lesssim
  0.5$. As a consequence, such a difference has a very small effect in
  our study---for example, on the order of $10^{-3}$ degrees for the
  range of parameters in Fig.~\ref{fig:chiPM}.  } Furthermore,
Ref.~\cite{Dlapa:2021vgp} argued that conservative memory terms are
still missing at 4PM order. However, at the PM order we are considering, there is 
no unique definition of the conservative dynamics. In this paper, we follow the
definition of the conservative dynamics of Refs.~\cite{Bini:2021gat,Bini:2020hmy,Bini:2020nsb}, which 
implies that memory effects at 4PM order will appear in the dissipative dynamics, and will be accounted for 
when they become available. Thus, here we assume that the conservative-dynamics results in
Refs.~\cite{Bern:2021dqo,Bern:2021yeh,Dlapa:2021npj,Dlapa:2021vgp} are
complete.

The two-body dynamics can be conveniently encoded in the gauge-invariant radial action, $I_r$, which at 4PM order can be written schematically as
\begin{align}
\label{Ir}
I_{r,\text{4PM}}^\text{hyp} &= I_{r,\text{3PM}} - \frac{\pi  G^4 M^7  \nu^2 P^2}{8 E L^3} \Bigg[\mathcal{M}_4^\text{p} \nonumber\\
&
+ \nu\left(4 \mathcal{M}_4^\text{t} \ln\frac{\sqrt{\gamma^2-1}}{2} +\mathcal{M}_4^{\pi^2} + \mathcal{M}_4^\text{rem}\right)\Bigg],
\end{align}
which includes the lower PM orders, with the 3PM part $I_{r,\text{3PM}}$ valid for both bound and unbound motion.
The terms $\mathcal{M}_4^{\dots}$ are directly related to parts of the scattering amplitude; they are independent of the masses, and are written in Eq.~(3) of Ref.~\cite{Bern:2021yeh}.
An expression for these terms that is valid for generic orbits (bound and unbound) is difficult to derive and has not yet been found.
The physical reason is that the tail effects~\cite{Blanchet:1987wq} start to enter at 4PM order, which is a nonlocal-in-time interaction depending on the entire history of the binary. Thus, it is different for bound and unbound orbits.

The scattering angle, by which the two bodies are deflected in the center-of-mass frame, is a gauge-invariant function that contains the same information as the radial action or the hyperbolic-orbit Hamiltonian. It can be obtained from the derivative of the radial action with respect to the angular momentum, that is, 
\begin{equation}
\label{chiPM}
\chi = -\frac{\partial I_r^\text{hyp}}{\partial L}-\pi.
\end{equation}

The 3PM and 4PM pieces of the conservative scattering angle have a logarithmic divergence in the high-energy (massless) limit. 
However, that divergence at 3PM order was shown to cancel with a corresponding divergence in the radiative contribution~\cite{Damour:2020tta,DiVecchia:2020ymx,DiVecchia:2021bdo,DiVecchia:2021ndb,Herrmann:2021tct,Bjerrum-Bohr:2021din}, and it is expected that such divergence also cancels at 4PM order~\cite{Bini:2021gat}.
In this paper, we only consider comparable-mass binaries, for which the singularity in the massless limit is irrelevant, and we demonstrate in Fig.~\ref{fig:chiPM} that the radiative contribution to the 3PM scattering angle is negligible for the range of parameters we consider.

Reference~\cite{Bern:2021yeh} also derived a two-body Hamiltonian from the radial action, following the steps in Refs.~\cite{Cheung:2018wkq,Bern:2019crd}.
The 4PM Hamiltonian in \emph{isotropic} gauge, and for hyperbolic orbits, is given by
\begin{equation}
\label{H4PM}
H_\text{4PM}^\text{hyp} = \sqrt{m_1^2 + P^2} + \sqrt{m_2^2 + P^2} + \sum_{n=1}^{4} \frac{G^n}{R^n} c_n,
\end{equation}
where the $c_n$ coefficients are given by Eqs.~(10) of Ref.~\cite{Bern:2019nnu} and Eq.~(8) of Ref.~\cite{Bern:2021yeh}.
Like the radial action, the 3PM part of $H_\text{4PM}^\text{hyp}$ is valid for generic motion, but the 4PM piece is for hyperbolic orbits. 
This Hamiltonian is determined in Ref.~\cite{Bern:2021yeh} from an ansatz that matches the scattering angle that follows from the radial action $I_{r,\text{4PM}}^\text{hyp}$, which is determined from the scattering amplitude.

To assess how close $H_\text{4PM}^\text{hyp}$ is to a bound-orbit 4PM Hamiltonian, we complement $H_\text{4PM}^\text{hyp}$ with bound-orbit corrections $\Delta H_\text{4PM(nPN)}^\text{ell}$, such that the $n$PN expansion of $H_\text{4PM}^\text{hyp}+\Delta H_\text{4PM(nPN)}^\text{ell}$ gives the correct $n$PN Hamiltonian up to $\Order(G^4)$ for bound orbits in isotropic gauge. 
We obtain $\Delta H_\text{4PM(nPN)}^\text{ell}$ to 6PN order, as explained in Appendix~\ref{app:Hiso}, since the 6PN Hamiltonian is fully known up to $\Order(G^4)$~\cite{Bini:2020hmy,Bini:2020nsb}.
In Sec.~\ref{sec:energy}, we compare the binding energy calculated from these Hamiltonians with NR, finding a small difference between $H_\text{4PM}^\text{hyp}$ and its bound-orbit corrections.

\section{Effective-one-body Hamiltonians}
\label{sec:EOBHam}

In the case of nonspinning compact objects, the EOB formalism~\cite{Buonanno:1998gg,Buonanno:2000ef} maps the binary motion to that of a test mass in a deformed Schwarzschild background. The two-body Hamiltonian $H^\text{EOB}$ is related to an effective Hamiltonian $H^\text{eff}$ via the energy map
\begin{equation}
H^\text{EOB} = M \sqrt{1 + 2 \nu \left(\frac{H_\text{eff}}{\mu} - 1\right)}\,.
\end{equation}
The effective metric is defined by
\begin{equation}
g_{\mu\nu}^\text{eff} \mr dx^\mu \mr dx^\nu = -A \mr dT^2 + B \mr dR^2 +  R^2 \left(\mr d\theta^2 + \sin^2\theta \mr d\phi^2\right),
\end{equation}
with the mass-shell condition~\cite{Damour:2000we}
\begin{equation}
0 = g^{\mu\nu}_\text{eff} P_\mu P_\nu + \mu^2 + Q,
\end{equation}
leading to the effective Hamiltonian $H_\text{eff} = - P_0$,
\begin{equation}
\label{effHamPots}
H_\text{eff}^2 = A \left(\mu^2 + \frac{P_R^2}{B} + \frac{L^2}{R^2} + Q\right).
\end{equation}
In the  $\nu \to 0$ limit, $H_\text{eff}$ reduces to the Schwarzschild Hamiltonian for a test mass, which is given by
\begin{equation}
\hat{H}_S^2 = (1-2u) \left[1 + (1-2u) p_r^2 + l^2u^2\right].
\end{equation}

To include higher PN information in an EOB Hamiltonian, we write an ansatz for the $A,B$ and $Q$ potentials in Eq.~\eqref{effHamPots}, perform a canonical transformation, then match $H^\text{EOB}$ to the PN-expanded Hamiltonian. This procedure is explained in more detail in Appendix~\ref{app:Hiso}.
For PM results, it is more convenient to calculate the gauge-invariant scattering angle from the ansatz for $H^\text{EOB}$, then match it to the PM-expanded scattering angle in Eq.~\eqref{chiPM}. These 
calculations will lead to the first derivation of 4PM-EOB Hamiltonians (and their $n$PN limits) for hyperbolic orbits. 

As an ansatz, we choose the $B$ potential to be the same as in the Schwarzschild metric, i.e., $B = 1/(1-2u)$, and include all the 4PM corrections into either $Q$ or $A$.
When included in $Q$, we get a 4PM generalization of the \emph{post-Schwarzschild} (PS) Hamiltonian considered in Refs.~\cite{Damour:2017zjx,Antonelli:2019ytb}, which is given by
\begin{align}
\label{HeffPS}
&(\hat{H}^\text{eff,PS})^2 = \hat{H}_S^2 + (1 - 2 u) \nonumber\\
&\qquad\times \left(u^2 q_\text{2PM} + u^3 q_\text{3PM} + u^4 q_\text{4PM}
+ \Delta_\text{4PN}^Q\right),
\end{align}
where $q_{n\text{PM}}$ can in general be any scalar function of the energy or the dynamical variables. In this ansatz, we include a 4PN correction term $\Delta_\text{4PN}^Q$ as explained below.

The other gauge we consider in this paper incorporates PM corrections in the $A$ potential, and reads
\begin{align}
\label{HeffPStild}
(\hat{H}^{\text{eff},\text{PS}^*})^2 &= \big(1-2u + u^2 a_\text{2PM} + u^3 a_\text{3PM} + u^4 a_\text{4PM} \nonumber\\
&\qquad
+ \Delta_\text{4PN}^A\big) \left[1 + (1-2u) p_r^2 + l^2u^2\right],
\end{align}
which is meant to more closely resemble, in the circular-orbit limit, the standard EOB gauge of Refs.~\cite{Buonanno:1998gg,Damour:2000we,Damour:2015isa}, which is often used in EOB waveform models for LIGO-Virgo-KAGRA observations. 

To determine the $q_{n\text{PM}}$ and $a_{n\text{PM}}$ coefficients, we calculate the scattering angle from the Hamiltonian.
To achieve this, we invert the Hamiltonian $\hat{H}^\text{EOB}(l, r, p_r)=\bar{E}+1/\nu$ to obtain $p_r(\bar{E},l,r)$, then evaluate the integral
\begin{equation}
\label{chiHam}
\chi= -2 \int_{r_0}^{\infty}\frac{\partial p_r(\bar{E},l,r)}{\partial l}dr - \pi\,,
\end{equation}
where $r_0$ is the turning point obtained from the largest root of the unperturbed (1PM) radial momentum $p_r^{(0)}(\bar{E},l,r)=0$.
To simplify evaluating this integral, we assume that a canonical transformation has been performed such that $q_{n\text{PM}}$ and $a_{n\text{PM}}$ are functions only of the effective energy $E_\text{eff}$, which is constant. 
However, since $q_{n\text{PM}}$ and $a_{n\text{PM}}$ themselves define $H_\text{eff}$, their dependence on $E_\text{eff}$ should be understood perturbatively in the PM scheme.
When working up to 3PM order, that energy can be taken to be the (1PM-accurate) Schwarzschild Hamiltonian $\hat{H}_S$ (see Refs.~\cite{Damour:2017zjx,Antonelli:2019ytb} for more details). However, at 4PM order, we need to account for nonlinear effects by using the 2PM effective energy, as explained in Appendix~\ref{app:eobcoefs}.

Since the scattering angle is gauge invariant, matching $\chi$ calculated from the EOB Hamiltonian to the PM-expanded scattering angle in Eq.~\eqref{chiPM} enables us to solve for the coefficients $q_{n\text{PM}}(\gamma)$ and $a_{n\text{PM}}(\gamma)$, where $\gamma\equiv E_\text{eff}/\mu$. (See Appendix~\ref{app:eobcoefs} or the Supplemental Material~\cite{ancmaterial} for the expressions of these coefficients.). 

\begin{table}[t]
\caption{Summary of the Hamiltonians considered in this paper.}
\begin{ruledtabular}
\begin{tabular}{p{0.35\linewidth} p{0.63\linewidth}}
Hamiltonian  &  Definition \\ 
\hline 
$H_{n\text{PN}}$ & PN-expanded Hamiltonian to $\Order(c^{-2n})$  \\
$H_{n\text{PM}}$ & PM-expanded Hamiltonian to $\Order(G^n)$  \\
$H_\text{4PM}^\text{hyp}+\Delta H_{\text{4PM}(n\text{PN})}^\text{ell}$ & 4PM hyperbolic-orbit Hamiltonian plus a bound-orbit correction up to orders $n$PN and 4PM \\
$H_\text{nPN(4PM)}^\text{ell,iso}$ & PN-expanded Hamiltonian truncated at $\Order(G^4)$ in isotropic coordinates, valid for bound orbits \\
$H_{n\text{PN}}^\text{EOB}$ & PN-EOB Hamiltonian in the gauge used in Refs.~\cite{Damour:2015isa,Buonanno:1998gg,Damour:2000we}\\
$H_{\dots}^\text{EOB,PS}$ & EOB Hamiltonian in Eq.~\eqref{HeffPS}, based on the PS gauge~\cite{Damour:2017zjx,Antonelli:2019ytb}  \\
$H_{\dots}^{\text{EOB},\text{PS}^*}$ & EOB Hamiltonian in the gauge used in Eq.~\eqref{HeffPStild} \\
$H_{\text{4PM}+\text{4PN}}^\text{EOB,\dots}$ & 4PM-EOB Hamiltonian (for hyperbolic orbits) complemented with the missing 4PN part (for bound orbits). \\
\end{tabular}
\end{ruledtabular}
\label{tab:Hdefs}
\end{table}

We also complement the 4PM EOB Hamiltonians above with the missing 4PN(5PM) piece. This is done by writing an ansatz for $\Delta_\text{4PN}$, such that
\begin{align}
\label{anz4PN}
\Delta_\text{4PN} &= \sum_{n=2}^{5} \alpha_n u^n \varepsilon^{5-n} 
+ \left(\alpha_\text{4,ln} u^4 \varepsilon + \alpha_\text{5,ln} u^5\right) \ln u \nonumber\\
&\quad 
+ \alpha_{\text{4,ln}\varepsilon} u^4 \ln \varepsilon,
\end{align}
where we use the PN expansion parameters $u$ and $\varepsilon \equiv \gamma^2 - 1$, each of one PN order.
Then, we perform a canonical transformation and match the result to the elliptic 4PN Hamiltonian of Ref.~\cite{Damour:2015isa}.
Note that the $\ln \varepsilon$ term is there to cancel a corresponding term that appears in the 4PM hyperbolic-orbit Hamiltonian.
The coefficients in Eq.~\eqref{anz4PN} are written in Appendix~\ref{app:eobcoefs}.

We summarize in Table~\ref{tab:Hdefs} the different Hamiltonians considered in this paper.

\section{Binding energy for circular orbits}
\label{sec:energy}
The 4PM part of the Hamiltonian in Eq.~\eqref{H4PM} is valid in the large-eccentricity limit, which means it is not consistent with the circular-orbit binding energy at 4PN and higher orders.
However, we show that the tail contribution to that Hamiltonian has a small effect on the dynamics, and so we use it to get an estimate for the 4PM contribution to the binding energy.

\subsection{PN-expanded binding energy}
We start by comparing the PN-expanded binding energy calculated from a bound-orbit Hamiltonian to the unbound case. 

In Appendix~\ref{app:Hiso}, we compute the bound-orbit 6PN Hamiltonian in isotropic coordinates, by canonically transforming the EOB Hamiltonian of Refs.~\cite{Bini:2020hmy,Bini:2020nsb}. We truncate that Hamiltonian at 4PM, and calculate the binding energy, which at 4PN reads
\begin{align}
\label{E4PNell}
\bar{E}_\text{4PN(4PM)}^\text{iso,ell} &= \bar{E}_\text{3PN}(x) + 
x^5 \bigg[\frac{11795}{768}-\nu \frac{518}{45} \ln x \nonumber\\
&\quad
+\nu  \left(\frac{428071 \pi ^2}{36864}-\frac{7899659}{34560}+\frac{1036 \ln 2}{45}\right) \nonumber\\
&\quad
+\nu ^2 \left(\frac{1435 \pi ^2}{576}-\frac{122815}{6912}\right) -\frac{5341 \nu ^3}{3456} \nonumber\\
&\quad
-\frac{77 \nu ^4}{62208}\bigg],
\end{align}
where $\bar{E}_\text{3PN}(x)$ is given by Eq.~(232) of Ref.~\cite{Blanchet:2013haa}, and $x\equiv (M\Omega)^{2/3}$, with $\Omega$ being the orbital frequency.
Note that the 4PN part of this result is \emph{not} gauge invariant, but is only valid for isotropic coordinates.
The reason is that PN-accurate coordinate transformations---for example to isotropic coordinates---in general span several PM orders.

We find that the difference between the binding energy in Eq.~\eqref{E4PNell} and the energy computed from the hyperbolic-orbit Hamiltonian~\eqref{H4PM} at 4PN is given by
\begin{align}
&\bar{E}_\text{4PN(4PM)}^\text{iso,hyp} - \bar{E}_\text{4PN(4PM)}^\text{iso,ell} \nonumber\\
&\qquad
= \nu  x^5 \bigg[\frac{37933}{45}+\frac{1036 \gamma_E}{45} -\frac{113847608 \ln 2}{45}\nonumber\\
&\quad\qquad
-\frac{1472499 \ln 3}{20}+\frac{13671875 \ln 5}{12}\bigg] \nonumber\\
&\qquad \simeq 14.94 \nu  x^5\,,
\end{align}
where we see that the disagreement is in the non-logarithmic, linear-in-$\nu$ coefficient of Eq.~\eqref{E4PNell}.
That coefficient is $-112.96$ in $\bar{E}_\text{4PN(4PM)}^\text{iso,ell}$, and is $-98.01$ in $\bar{E}_\text{4PN(4PM)}^\text{iso,hyp}$, with the difference being $14.94$.
(Note that the $\ln x$ term in Eq.~\eqref{E4PNell} is the same for bound and unbound orbits, as was shown to all PN orders in Ref.~\cite{Cho:2021arx}.)

\begin{figure}[t]
\centering
\includegraphics[width=\linewidth]{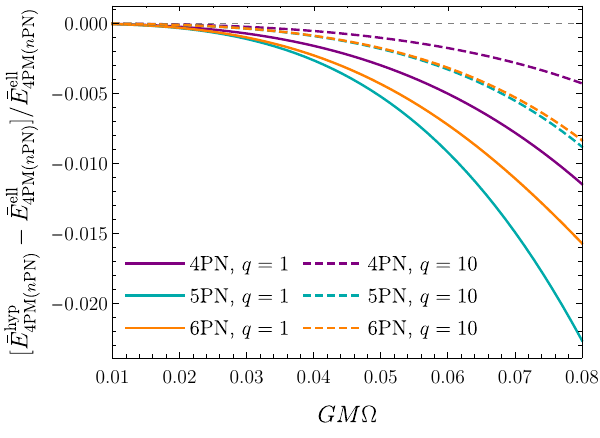}
\caption{Relative difference in the circular-orbit binding energy, computed analytically from an elliptic versus a hyperbolic-orbit Hamiltonian.}
\label{fig:EbPN} 
\end{figure}

In Fig.~\ref{fig:EbPN}, we plot the relative difference in the binding energy at different PN orders, and for mass ratios $q=1$ and $q = 10$, finding disagreement $\lesssim 2\%$, which justifies applying the hyperbolic 4PM result to bound orbits as we do below, since we find that the disagreement with NR is larger than $2\%$.

\subsection{Binding energy from PM-expanded Hamiltonians}

\begin{figure*}[t]
\centering
\includegraphics[width=0.49\linewidth]{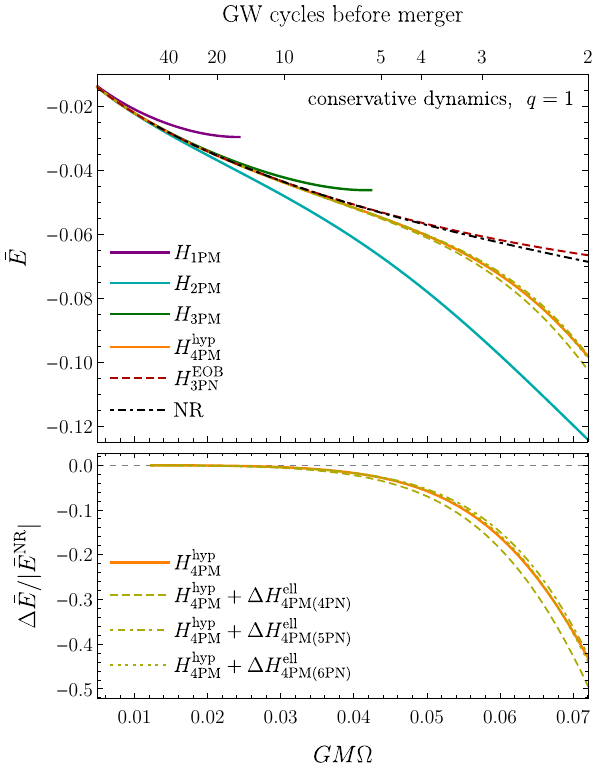}
\includegraphics[width=0.49\linewidth]{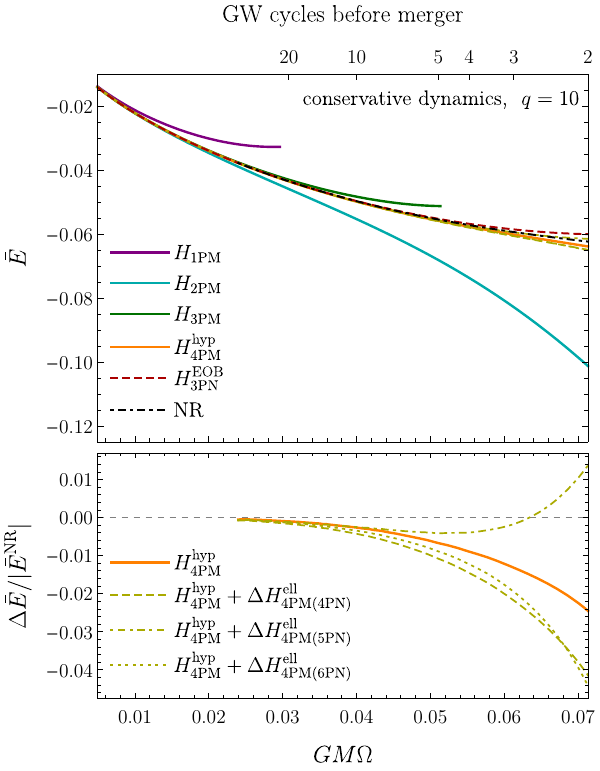}
\caption{Binding energy versus orbital frequency for the PM-expanded Hamiltonians compared to the NR prediction for a nonspinning equal-mass (left panel) and mass-ratio $q=10$ (right panel) binary black hole. The top axis gives the number of GW cycles before merger, which is twice the number of orbits. The lower panels show the relative difference of the 4PM curves with NR.}
\label{fig:EbPM} 
\end{figure*}

To compute the binding energy for circular orbits from the PM-expanded Hamiltonian in Eq.~\eqref{H4PM}, we do so numerically by setting $p_r=0$ in the Hamiltonian and solving $\dot{p}_r=-\partial \hat{H}/\partial r=0$ for the angular momentum $l$ at different orbital separations.
We then plot $\bar{E} = \hat{H}_{n\text{PM}} - 1/\nu$ versus the orbital frequency $M\Omega = \partial \hat{H}/ \partial l$ (see Ref.~\cite{Antonelli:2019ytb} for more details).

In Fig.~\ref{fig:EbPM}, we compare the binding energy with NR data that were extracted in Ref.~\cite{Ossokine:2017dge} from numerical simulations produced by the Simulating eXtreme Spacetimes (SXS) collaboration~\cite{SXS,Boyle:2019kee}. In particular, we use the simulations with SXS ID 0180 for mass ratio $q=1$ and ID 0303 for $q=10$, for which the numerical error is too small to show in the figure.

We see that the 4PM Hamiltonian gives much better agreement with NR toward merger than Hamiltonians computed at lower PM orders. This is because the 4PM Hamiltonian contains the full 3PN information, which is known to give considerably better results than 2PN.
The improvement at 4PM is even more significant for mass ratio $q=10$ than $q=1$, because the 3PN coefficient in the binding energy increases significantly with increasing mass ratio. 
In addition, it seems that $H_\text{4PM}^\text{hyp}$ is very close to what a bound-orbit 4PM Hamiltonian would be, as evidenced by how close the curves $H_\text{4PM}^\text{hyp}+\Delta H_\text{4PM(nPN)}^\text{ell}$, computed at 4PN, 5PN and 6PN orders, are to $H_\text{4PM}^\text{hyp}$.

For comparison, the figure also shows the 3PN EOB Hamiltonian in the gauge of Refs.~\cite{Buonanno:1998gg,Damour:2000we,Damour:2015isa}, which gives better agreement with NR (also when considering 
different mass ratios) because it includes the exact test-body limit, and the associated resummation of the PN results.
In the plots, we stop the numerical evaluation of the energy either at the innermost-stable circular orbit (ISCO) or at two GW cycles (one orbit) before merger. 
We note that the EOB results on the figures do not contain any NR information, and the $g_{00}$ effective metric contains PN terms in a Taylor-expanded form.

\subsection{Binding energy from PM-EOB Hamiltonians}

\begin{figure*}
\centering
\vspace{-0.5cm}
\includegraphics[width=0.49\linewidth]{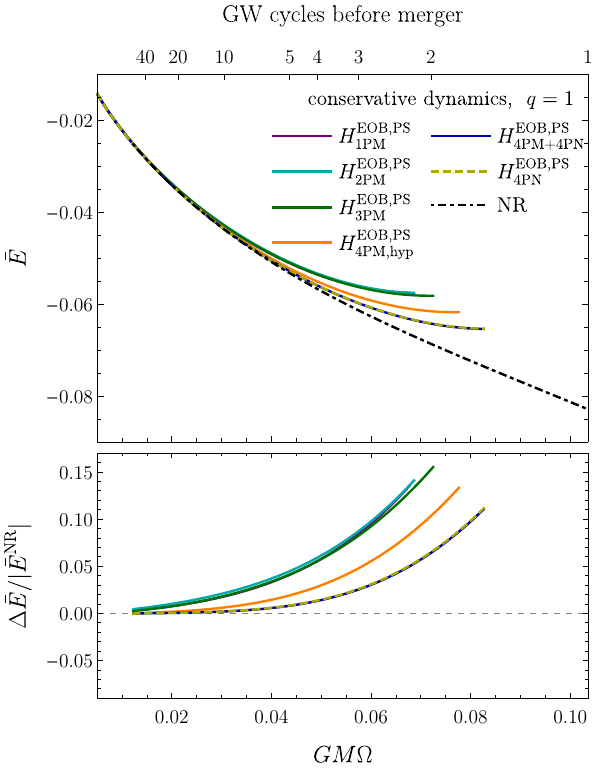}
\includegraphics[width=0.49\linewidth]{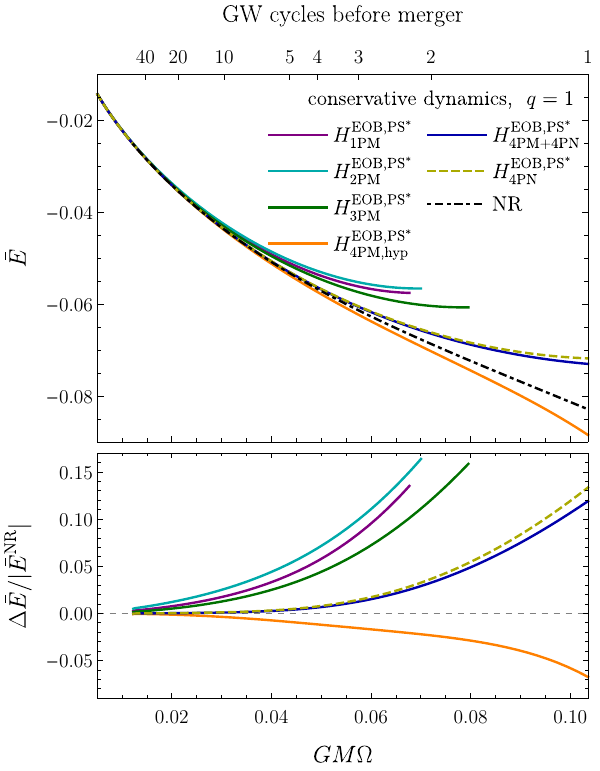}
\caption{Binding energy versus orbital frequency for the PM-EOB Hamiltonians, compared to the NR prediction for $q = 1$. The left panels contain results for the EOB Hamiltonian in the PS gauge in Eq.~\eqref{HeffPS}, while the right panels are for the \PSstr gauge in Eq.~\eqref{HeffPStild}.}
\label{fig:EbEOB}
\includegraphics[width=0.49\linewidth]{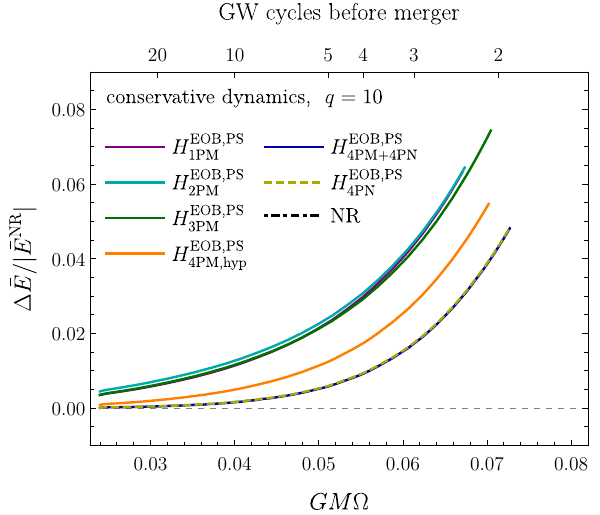}
\includegraphics[width=0.49\linewidth]{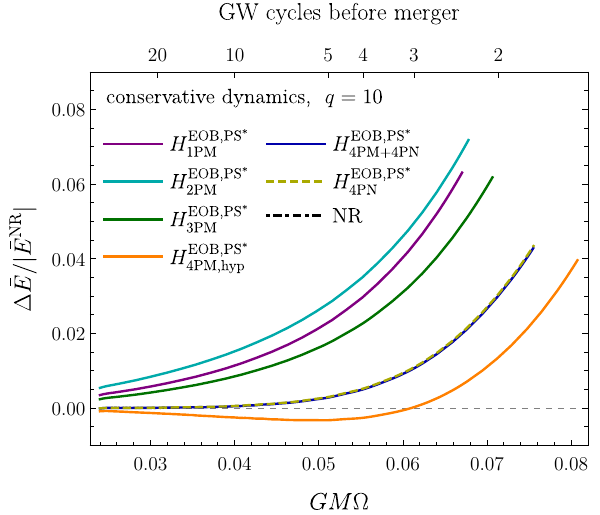}
\caption{Similar to Fig.~\ref{fig:EbEOB} but for mass ratio $q = 10$, and we only show the relative difference since the EOB curves are closer to the NR curve than for equal masses. All curves end at the ISCO.}
\label{fig:EbEOB10}
\end{figure*}
\begin{figure*}[t]
\centering
\includegraphics[width=0.48\linewidth]{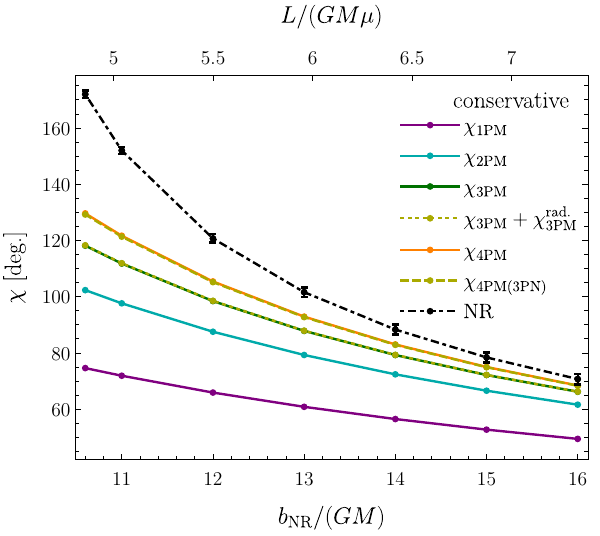}
\includegraphics[width=0.48\linewidth]{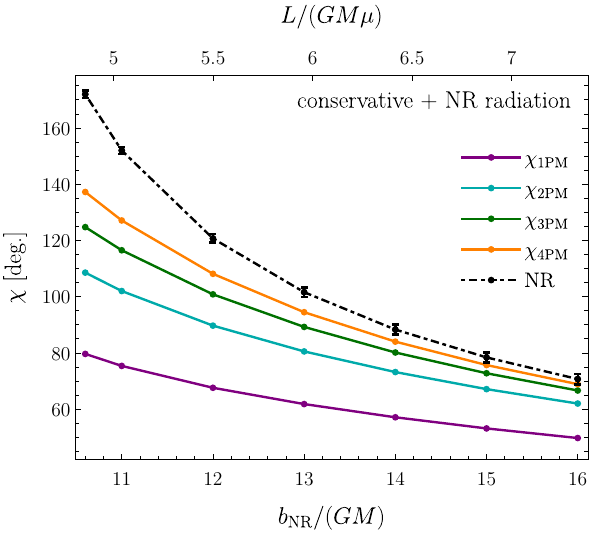}
\caption{Comparison of the PM-expanded scattering angle with NR. The left panel shows the conservative scattering angle, except for the $\chi_\text{3PM}+\chi_\text{3PM}^\text{rad.}$ curve that also includes the leading-order (3PM) radiative contribution. The right panel incorporates the effect of the radiative losses from the NR simulations using Eq.~\eqref{chiTot}.}
\label{fig:chiPM}
\end{figure*}

Similarly, computing the binding energy from the PM-EOB Hamiltonians leads to Fig.~\ref{fig:EbEOB} for $q=1$ and Fig.~\ref{fig:EbEOB10} for $q = 10$.
In both figures, we plot the binding energy at each PM order, and complement 4PM with 4PN information. We stop the numerical evaluation either at the ISCO or at one GW cycle before merger.

From the figures, we observe the following:
\begin{itemize}
\item The 4PM order provides a significant improvement over the lower orders, even though the PM Hamiltonian is for hyperbolic orbits.
\item The \PSstr gauge performs better than the PS gauge, for both equal and unequal masses. That difference is due to higher-order terms resulting from the resummation of the Hamiltonian coefficients. We expect that the more PN/PM orders are included, the closer different gauges would be to each other.
\item Complementing 4PM with the missing 4PN information for bound orbits gives results for the \PSstr gauge that are comparable to the standard PN-EOB gauge.
\item Using 4PN-expanded potentials in $H_\text{4PN}^\text{EOB,...}$ gives almost the same result as the 4PM+4PN Hamiltonian, although for the \PSstr gauge, the 4PM+4PN Hamiltonian is slightly better for equal masses than its 4PN expansion.
\end{itemize}

\section{Scattering angle comparison with NR}
\label{sec:scattering}

Since the 4PM part of the radial action in Eq.~\eqref{Ir} is valid for hyperbolic orbits, a better comparison with NR is through the scattering angle.

NR simulations for the scattering angle were reported in Ref.~\cite{Damour:2014afa} for equal masses $q = 1$ and initial linear momentum $|\bm{p}| = 0.11456439 M$.
The initial energy in these simulations is approximately $E_\text{in}^\text{NR}\simeq 1.02259$ (corresponding to velocity $v \simeq 0.4$), and the initial angular momentum $L_\text{in}^\text{NR}$ is proportional to the impact parameter $b_\text{NR}$, which ranges between 9$M$ and 16$M$. 
The NR error in the scattering angle is $\sim 1 - 2$ degrees.

Reference~\cite{Damour:2014afa} also reported the energy and angular momentum losses due to the emitted GWs, which can be used to account for the radiative contribution to the scattering angle. 
It was proven in Ref.~\cite{Bini:2012ji} that when working to linear order in RR, the radiative contribution to the total scattering angle is half the difference of the conservative scattering angle evaluated as a function of the outgoing and incoming states, i.e.,
\begin{equation}
\chi^\text{rad.} = \frac{1}{2} \left[\chi^\text{cons.}(E_\text{out}, L_\text{out}) - \chi^\text{cons.}(E_\text{in}, L_\text{in})\right],
\end{equation}
which means that the total scattering angle is given by
\begin{align}
&\chi^\text{tot}(E_\text{in}, L_\text{in}) \equiv \chi^\text{cons.}(E_\text{in}, L_\text{in}) + \chi^\text{rad.}(E_\text{in}, L_\text{in}) \nonumber\\
&\qquad
=\frac{1}{2} \left[\chi^\text{cons.}(E_\text{in}, L_\text{in}) + \chi^\text{cons.}(E_\text{out}, L_\text{out})\right],
\end{align}
which can be written as
\begin{equation}
\label{chiTot}
\chi^\text{tot}(E_\text{in}, L_\text{in}) = \chi^\text{cons.}(E_\text{avg}, L_\text{avg}),
\end{equation}
where 
\begin{equation}
E_\text{avg} \equiv \frac{E_\text{in} + E_\text{out}}{2}, \quad
L_\text{avg} \equiv \frac{L_\text{in} + L_\text{out}}{2}.
\end{equation}
We emphasize that Eq.~\eqref{chiTot} holds when neglecting contributions quadratic in RR, which start at 5PN order.

\subsection{PM-expanded scattering angle}

In the left panel of Fig.~\ref{fig:chiPM}, we plot the conservative PM-expanded scattering angle, calculated from Eq.~\eqref{chiPM}, for the initial values of energy and angular momentum used in the NR simulations, which are written in Table I of Ref.~\cite{Damour:2014afa}. 
We see that each PM order gives better agreement with NR than the lower orders, with an overall good agreement at 4PM, especially considering that these scattering angles are rather large while the PM expansion is an approximation away from a straight line.

We also plot the 3PN expansion of the 4PM scattering angle, finding that the difference with the full 4PM angle is only $\sim .1$ degrees. This is because the initial velocity is $v\simeq 0.4$, for which a PN expansion provides a good approximation. Higher velocities would lead to larger differences between the PM scattering angle and its PN expansion.
Furthermore, the figure shows the 3PM conservative scattering angle plus the leading order (3PM) radiative contribution, which is given by Eq.~(5.7) of Ref.~\cite{Damour:2020tta}; however, the effect of that radiative contribution is so small that it is almost the same as the conservative 3PM curve.

In the right panel of Fig.~\ref{fig:chiPM}, we plot the scattering angle taking into account the effect of RR by modifying the initial conditions, as explained above in Eq.~\eqref{chiTot}.  
We see that all curves are shifted closer to the NR curve, but their order remains the same.

\subsection{Scattering angle from PM-EOB Hamiltonians}

\begin{figure*}[t]
\includegraphics[width=0.48\linewidth]{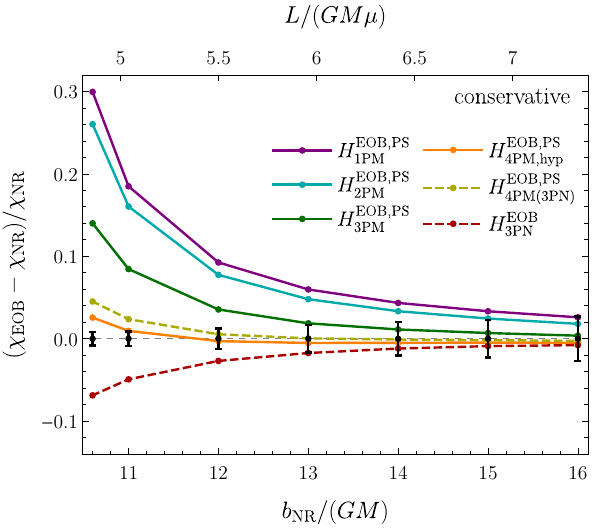}
\includegraphics[width=0.48\linewidth]{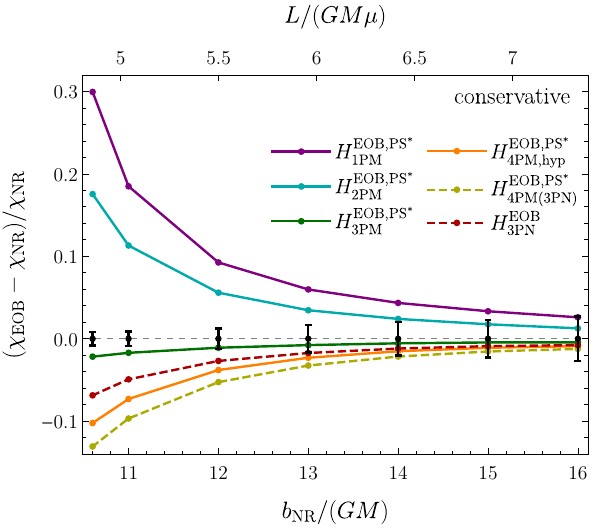}
\caption{Comparison with NR of the conservative scattering angle computed from the EOB Hamiltonians in the PS gauge (left panel) and the \PSstr gauge (right panel).}
\label{fig:chiEOB}
\vspace{\floatsep}
\includegraphics[width=0.48\linewidth]{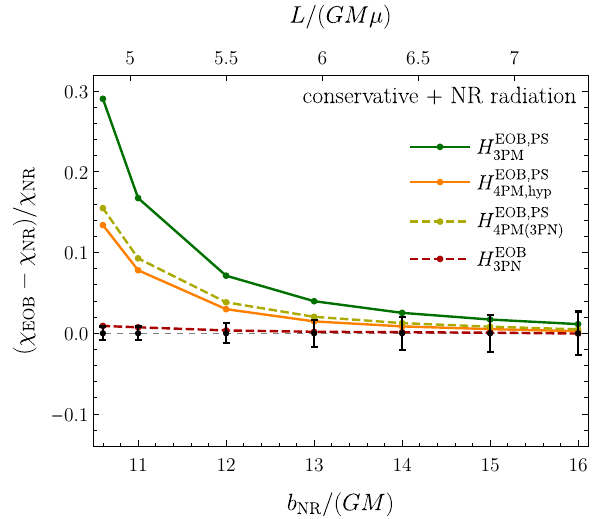}
\includegraphics[width=0.48\linewidth]{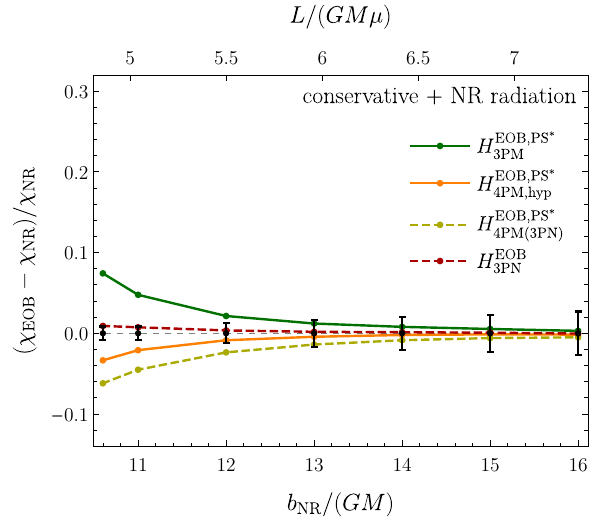}
\caption{Scattering angle calculated from the EOB Hamiltonians in the PS gauge (left panel) and the \PSstr gauge (right panel) while incorporating radiative effects from NR through the initial conditions as in Eq.~\eqref{chiTot}.}
\label{fig:chiEOBRR}
\end{figure*}

To compute the scattering angle from the EOB Hamiltonians, we start with an initial azimuthal angle $\phi_\text{in}=0$, evolve the equations of motion without RR force, then read off the final angle $\phi_\text{out}$, leading to the scattering angle $\chi_\text{EOB} = \phi_\text{out} - \phi_\text{in} - \pi$.
We start the evolution with initial separation $r_\text{in}= 10^4$, initial angular momentum $L_\text{in}^\text{NR}$, and solve $E_\text{in}^\text{NR} = H^\text{EOB}$ for the initial $p_r$.

In Fig.~\ref{fig:chiEOB}, we plot the relative difference in the conservative scattering angle between EOB and NR, finding much smaller difference than the PM-expanded angles in Fig.~\ref{fig:chiPM}.
Similarly, in Fig.~\ref{fig:chiEOBRR}, we plot the same quantity but using initial conditions that account for the NR energy and angular momentum losses, as in Eq.~\eqref{chiTot}.
For comparison, both figures show the scattering angle calculated from a 3PN EOB Hamiltonian in the gauge of Refs.~\cite{Damour:2015isa,Buonanno:1998gg,Damour:2000we}, which is valid for arbitrary trajectories since 3PN is purely local in time (no tails contributions). 

From Fig.~\ref{fig:chiEOB}, we see that 4PM performs better than 3PM for the PS gauge, but 3PM is better for the \PSstr gauge. 
However, when including the NR RR in Fig.~\ref{fig:chiEOBRR}, 4PM becomes closer to NR than 3PM for both gauges. We also see that the \PSstr gauge gives better agreement with NR than the PS gauge.

For bound orbits, we saw in the previous section that PN-expanding the PM Hamiltonians gave almost the same results. 
For scattering encounters on the other hand, we see from both Fig.~\ref{fig:chiEOB} and Fig.~\ref{fig:chiEOBRR} that the 4PM Hamiltonians lead to better results than the 3PN expansion of their potentials, for both EOB gauges and whether or not RR is included.

\subsection{4PN EOB Hamiltonian for hyperbolic orbits}
The 4PN EOB Hamiltonian derived in Ref.~\cite{Damour:2015isa} included the tail part in a small-eccentricity expansion, and is thus valid for bound orbits.
The 4PN tail contribution for hyperbolic orbits was derived in Ref.~\cite{Bini:2017wfr}, analytically at leading order in the large-eccentricity expansion, and numerically for eccentricity $\geq 1$.
However, it is not straightforward to translate the all-orders-in-eccentricity result of Ref.~\cite{Bini:2017wfr} to an EOB Hamiltonian in the form of Eq.~\eqref{effHamPots}.
Therefore, here, we compute a hyperbolic-orbit 4PN EOB Hamiltonian at next-to-leading order in the large-eccentricity expansion.
We use that Hamiltonian to compare the effect of using an elliptic versus a hyperbolic-orbit Hamiltonian on the scattering angle.

We start by writing an effective Hamiltonian of the form
\begin{align}
\label{H4PNhyp}
\hat{H}_\text{eff}^\text{4PN,hyp} &= \sqrt{A^\text{hyp} \left[1 + \frac{p_r^2}{B^\text{hyp}} + l^2 u^2 + Q^\text{hyp}\right]}\,, \\
(B^\text{hyp})^{-1} &\equiv A^\text{hyp} \bar{D}^\text{hyp},
\end{align}
in which the potentials contain local and nonlocal-in-time contributions, starting at 4PN order. 
The local part is valid for generic motion, and is given by Eqs. (4.4), (9.6) and (9.7) of Ref.~\cite{Bini:2020wpo}, which we follow in how the local and nonlocal parts are split.

We then write an ansatz for the potentials with unknown coefficients for the nonlocal contribution, that is
\begin{align}
\label{AhypAnz}
&A^\text{hyp} = 1 - 2 u + 2\nu u^3 + \left(\frac{94}{3} - \frac{41}{32} \pi^2\right) \nu u^4 \nonumber\\
&\qquad 
+ \left[\left(\frac{41 \pi ^2}{32}-\frac{221}{6}\right) \nu ^2+\left(\frac{2275 \pi ^2}{512}-\frac{4237}{60}\right) \nu\right] u^5 \nonumber\\
&\qquad + \left(a_5^\text{nloc} + a_{5,\ln u}^\text{nloc} \ln u + a_{5,\ln p}^\text{nloc} \ln p^2\right) u^5, 
\\
&\bar{D}^\text{hyp} = 1+6 \nu  u^2 + \left(52 \nu -6 \nu ^2\right) u^3 \nonumber\\
&\qquad 
+ \left[\left(\frac{123 \pi ^2}{16}-260\right) \nu ^2+\left(\frac{1679}{9}-\frac{23761 \pi ^2}{1536}\right) \nu\right] u^4 \nonumber\\
&\qquad+ \left(d_4^\text{nloc} + d_{4,\ln u}^\text{nloc} \ln u + d_{4,\ln p}^\text{nloc} \ln p^2\right) u^4, 
\\
&Q^\text{hyp} = 2 \nu  (4-3 \nu ) u^2 p_r^4 + \left(10 \nu ^3-83 \nu ^2+20 \nu\right) u^3 p_r^4 \nonumber\\
&\qquad
+ \left(6 \nu ^3-\frac{27 \nu ^2}{5}-\frac{9 \nu }{5}\right) u^2 p_r^6.
\end{align}
In this ansatz, the 4PM nonlocal coefficients in $\bar{D}^\text{hyp}$ are at leading order in the large-eccentricity expansion, while those at 5PM in $A^\text{hyp}$ are at next-to-leading order, with no nonlocal contributions to $Q^\text{hyp}$.
We also assume in the ansatz a dependence on $\ln p^2$ because it simplifies the result, but other possible choices include $\ln (l^2 u^2)$ or $\ln E_\text{N}$, with $E_\text{N}$ being the Newtonian energy.

To fix the unknown coefficients in the ansatz, we calculate the scattering angle from the Hamiltonian using Eq.~\eqref{chiHam}, then match the result to the \emph{total} 5PM(4PN) scattering angle, which schematically reads
\begin{equation}
\frac{\chi_\text{4PN}}{2} = \frac{\chi_1}{L} + \frac{\chi_2}{L^2} + \frac{\chi_3}{L^3} + \frac{\chi_4^\text{loc} + \chi_4^\text{nloc}}{L^4}  + \frac{\chi_5^\text{loc} + \chi_5^\text{nloc}}{L^5}\,,
\end{equation}
where the local $\chi_n$ coefficients are given by Eq.~(10.1) of Ref.~\cite{Bini:2020wpo}, and the nonlocal part by Eq.~(6.11) of Ref.~\cite{Bini:2020hmy}. After matching the scattering angle and solving for the Hamiltonian coefficients, we obtain the solution
\begin{align}
a_5^\text{nloc} &= \nu  \left(\frac{2752}{15} \ln 2-\frac{5464}{75}\right), \nonumber\\
a_{5,\ln p}^\text{nloc} &= \frac{64 \nu}{5}, \qquad
a_{5,\ln u}^\text{nloc} = 0, \nonumber\\
d_4^\text{nloc} &= \nu  \left(168-\frac{1184}{15} \ln 2\right),  \nonumber\\
d_{4,\ln p}^\text{nloc} &= \frac{592 \nu}{15}, \qquad
d_{4,\ln u}^\text{nloc} = 0, 
\end{align}
that is, with no dependence on $\ln u$.

\begin{figure}[t]
\centering
\includegraphics[width=\linewidth]{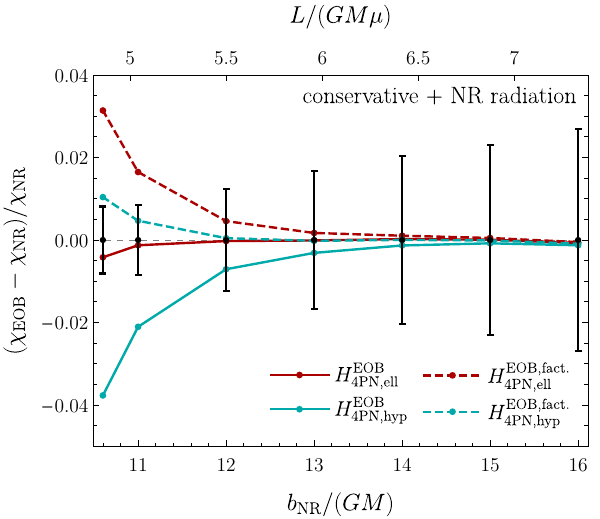}
\caption{Scattering angle calculated from 4PN EOB Hamiltonians for elliptic and hyperbolic orbits. The solid lines are for Hamiltonians with Taylor-expanded potentials, while the dashed lines are for Hamiltonians with factorized potentials.}
\label{fig:chi4PN} 
\end{figure}

In Fig.~\ref{fig:chi4PN}, we compare to NR the scattering angle computed from the elliptic-orbit Hamiltonian of Ref.~\cite{Damour:2015isa} and the angle computed from the hyperbolic-orbit Hamiltonian in Eq.~\eqref{H4PNhyp}.
We see that, unexpectedly, the elliptic-orbit Hamiltonian gives better agreement with NR.
However, that result depends on the particular resummation of the potentials. 

To illustrate this, we consider a simple factorization of the $A$ potential given by
\begin{align}
&A^\text{hyp,fact.} = \left(1 - 2u\right) \bigg\lbrace1 + 2 \nu u^3 + \left(\frac{106}{3}-\frac{41 \pi ^2}{32}\right)\nu u^4 \nonumber\\
&\qquad
+ \bigg[
\left(\frac{963 \pi ^2}{512}-\frac{21841}{300}+\frac{2752 \ln 2}{15}\right) \nu   \nonumber\\
&\quad\qquad
+ \left(\frac{41 \pi ^2}{32}-\frac{221}{6}\right) \nu ^2
+\frac{64}{5} \nu  \ln p^2
\bigg]u^5
\bigg\rbrace,
\end{align}
which agrees with Eq.~\eqref{AhypAnz} when Taylor expanded.
Similarly, the factorized version of the elliptic-orbits $A$ potential in Eq.~(8.1) of Ref.~\cite{Damour:2015isa} reads
\begin{align}
&A^\text{ell,fact.} = \left(1 - 2u\right) \bigg\lbrace1 + 2 \nu u^3 + \left(\frac{106}{3}-\frac{41 \pi ^2}{32}\right)\nu u^4 \nonumber\\
&\qquad
+ \bigg[
\left(\frac{1}{20}+\frac{963 \pi ^2}{512}+\frac{128 \gamma_E}{5}+\frac{256 \ln 2}{5}\right) \nu  \nonumber\\
&\quad\qquad
+\left(\frac{41 \pi ^2}{32}-\frac{221}{6}\right) \nu ^2 + \frac{64}{5} \nu\ln u
\bigg]u^5
\bigg\rbrace.
\end{align}
Comparing the scattering angle computed from Hamiltonians with these factorized potentials (dashed lines in Fig.~\ref{fig:chi4PN}), we see that the hyperbolic-orbit Hamiltonian now gives better agreement with NR than the one for elliptic orbits.

These results show that there can be differences between elliptic and hyperbolic-orbit Hamiltonians when applied to scattering encounters, but which performs better depends on the particular gauge of the Hamiltonian and the resummations of its coefficients.

\section{Conclusions}
\label{sec:conc}
In this paper, we investigated the conservative 4PM Hamiltonian with nonlocal-in-time (tail) effects for hyperbolic orbits, which was derived in 
Refs.~\cite{Bern:2021dqo,Bern:2021yeh,Dlapa:2021npj,Dlapa:2021vgp}, by comparing it to NR simulations for the binding energy and scattering angle.
We found an improvement over lower PM orders, which was expected since 4PM order contains the full 3PN information. 
In addition, even though the nonlocal part of the 4PM Hamiltonian is valid for hyperbolic motion, we showed that it performs well for bound orbits, and that the hyperbolic piece has a small effect on the dynamics. This was demonstrated by comparing the PN-expanded binding energy for bound versus unbound orbits (see Fig.~\ref{fig:EbPN}), and by complementing the 4PM Hamiltonian with bound-orbit corrections at 4PN, 5PN, and 6PN orders  (see Fig.~\ref{fig:EbPM}).

As a first study, we incorporated the 4PM information in two EOB Hamiltonians, given by Eqs.~\eqref{HeffPS} and \eqref{HeffPStild}.
(The EOB Hamiltonians are not calibrated to NR simulations, and 
do not use resummations of the effective-metric components.) For bound orbits, we found that the PM Hamiltonians gave similar results to the same Hamiltonians with PN-expanded potentials.
However, for the scattering angle, the PM-EOB Hamiltonians showed better agreement with NR than PN-EOB Hamiltonians in the same gauge (see Figs.~\ref{fig:chiEOB} and \ref{fig:chiEOBRR}).

In particular, we found that including 4PM results in EOB Hamiltonians improved the disagreement with the NR binding energy from about $40 \%$, for equal masses at two GW cycles before merger, to about $10\%$ for the PS gauge and $3\%$ for the \PSstr gauge (see Figs.~\ref{fig:EbPM} and \ref{fig:EbEOB}).
These results have been obtained for the conservative dynamics, but will change, and likely improve, once RR is included and the equations of motion are evolved for an inspiraling trajectory.
For the scattering angle, the differences with NR were $8\%$ and $2\%$, respectively for the two EOB gauges, at impact parameter $b = 11 GM$ and initial relative velocity $v \simeq 0.4$ (see Figs.~\ref{fig:chiPM} and \ref{fig:chiEOBRR}).
Our comparisons of the scattering angle also highlighted the importance of including RR effects even when comparing conservative results with NR. 
For example, the conclusions one draws would be different between Fig.~\ref{fig:chiEOB}, which is purely conservative, and Fig.~\ref{fig:chiEOBRR}, which includes the NR radiative losses.

Furthermore, we worked out a 4PN EOB Hamiltonian for hyperbolic orbits, which extends the elliptic-orbit Hamiltonian of Ref.~\cite{Damour:2015isa}. We compared the scattering angles of the two Hamiltonians to NR and showed that the Hamiltonian gauge and the resummations of its coefficients can affect the agreement with NR. 

The only NR simulations currently available in the literature for the scattering angle~\cite{Damour:2014afa} are for equal masses and for a specific value of the energy corresponding to $v\simeq 0.4$. 
It would be interesting to see how PM and PN information compare with NR for unequal masses and higher velocities. Such studies would enable the construction of accurate waveform models over the whole binary parameters space including large eccentricities and large velocities.

\section*{Acknowledgments} 
We are grateful to Zvi Bern, Enrico Herrmann, Radu Roiban, and Mikhail Solon for fruitful discussions and valuable comments.
We are also grateful to Gregor K\"{a}lin and Rafael Porto for discussions and for notifying us about a missing contribution to the 6PN isotropic-gauge Hamiltonian in an earlier version of the paper.
We thank Julio Parra-Martinez, Michael Ruf, Chia-Hsien Shen, and Mao Zeng for useful discussions, and thank Sergei Ossokine for providing the NR data for the binding energy.

\appendix

\section{PN Hamiltonian for bound orbits in isotropic gauge}
\label{app:Hiso}

In this Appendix, we canonically transform the 6PN-EOB Hamiltonian of Refs.~\cite{Bini:2020hmy,Bini:2020nsb} to the isotropic gauge, in which the Hamiltonian only depends on $r$ and $p^2$ with no explicit dependence on the angular momentum.

We start by writing an ansatz with unknown coefficients for the Hamiltonian
\begin{align}
\hat{H}_\text{6PN}^\text{iso} &= \frac{1}{\nu} + \frac{p^2}{2} - \frac{1}{r}
+ \sum_{i=2}^{7} \sum_{j=0}^{i} \alpha_{ij} \frac{p^{2(i-j)}}{r^j} \nonumber\\
&\quad
+\sum_{i=5}^{7} \sum_{j=1}^{i} \alpha_{ij} \frac{p^{2(i-j)}}{r^j} \ln r,
\end{align}
where the 0PM coefficients are given by the PN expansion of 
\begin{equation}
H_\text{0PM}^\text{iso} = \sqrt{m_1^2 + P^2} + \sqrt{m_2^2 + P^2}.
\end{equation}

Then, we write an ansatz for the generating function $\mathcal{G}$, perform a canonical transformation using Poisson brackets, and match to the 6PN EOB Hamiltonian, i.e.,
\begin{align}
H_\text{6PN}^\text{EOB} &= H_\text{6PN}^\text{iso} + \lbrace \mathcal{G}, H_\text{6PN}^\text{iso} \rbrace
+ \frac{1}{2!}\lbrace \mathcal{G},\lbrace \mathcal{G}, H_\text{6PN}^\text{iso} \rbrace\rbrace \nonumber\\
&\quad
+ \frac{1}{3!}\lbrace \mathcal{G},\lbrace \mathcal{G},\lbrace \mathcal{G}, H_\text{6PN}^\text{iso} \rbrace\rbrace\rbrace + \dots,
\end{align}
where each bracket introduces a factor of $1/c^2$.

The result for the full 6PN Hamiltonian, which contains 6 coefficients that have not yet been determined in Refs.~\cite{Bini:2020hmy,Bini:2020nsb}, is provided in the Supplemental Material. Here, we write that Hamiltonian truncated at $\Order(G^4)$
\begin{widetext}
\begin{align}
\label{Hiso}
&\hat{H}_\text{4PN(4PM)}^\text{ell,iso} = \hat{H}_\text{3PN}^\text{iso} 
+ \left[\frac{7}{256}+\frac{63 \nu ^4}{256}-\frac{105 \nu ^3}{128}+\frac{189 \nu ^2}{256}-\frac{63 \nu }{256}\right] p^{10} \nonumber\\
&\quad
+ \frac{G p^8}{r} \left[\frac{45}{128}-\nu ^4-3 \nu ^3+\frac{51 \nu ^2}{8}+\nu  \left(-\frac{294051}{2800}+\frac{10834496  \ln 2}{45}+\frac{6591861  \ln 3}{700}-\frac{27734375  \ln 5}{252}\right)\right] \nonumber\\
&\quad
+ \frac{G^2 p^6}{r^2}  \left[\frac{13}{8}+\frac{35 \nu ^4}{32}+\frac{337 \nu ^3}{16}+\frac{453 \nu ^2}{32}+\nu  \left(\frac{9062513}{16800}-\frac{21212984}{15}   \ln 2+\frac{1296484375  \ln 5}{2016}-\frac{282031389  \ln 3}{5600}\right)\right] \nonumber\\
&\quad
+ \frac{G^3 p^4}{r^3} \left[\frac{105}{32}-\frac{487 \nu ^3}{16}-\frac{2589 \nu ^2}{32}+\nu  \left(-\frac{2872367}{2400}+\frac{233388968  \ln 2}{75}+\frac{16351713  \ln 3}{160}-\frac{405859375  \ln 5}{288}\right)\right] \nonumber\\
&\quad
+\frac{G^4 p^2}{r^4} \bigg[\frac{105}{32}-\frac{5 \nu ^4}{16}+\frac{27 \nu ^3}{2}+\left(\frac{2957}{48}-\frac{41 \pi ^2}{64}\right) \nu ^2+\nu  \bigg(-\frac{74 \ln r}{15}+\frac{4086409}{3600}-\frac{29665 \pi ^2}{12288}+\frac{148 \gamma_E }{15} \nonumber\\
&\qquad
-\frac{680106004}{225}   \ln 2+\frac{196484375  \ln 5}{144}-\frac{37122381  \ln 3}{400}\bigg)\bigg], \\
&\hat{H}_\text{5PN(4PM)}^\text{ell,iso} = \hat{H}_\text{4PN(4PM)}^\text{ell,iso} 
+ \left[-\frac{21}{1024}+\frac{231 \nu ^5}{1024}-\frac{1155 \nu ^4}{1024}+\frac{1617 \nu ^3}{1024}-\frac{231 \nu ^2}{256}+\frac{231 \nu }{1024}\right] p^{12} \nonumber\\
&\quad
+ \frac{G p^{10}}{r} \bigg[-\frac{77}{256}-\nu ^5-\frac{5 \nu ^4}{2}+\frac{95 \nu ^3}{8}+\nu ^2 \left(-\frac{18236}{35}+\frac{10834496 \ln 2}{9}+\frac{6591861 \ln 3}{140}-\frac{138671875 \ln 5}{252}\right) \nonumber\\
&\qquad
+\nu  \left(\frac{1212079}{22400}-\frac{5417248}{45}  \ln 2+\frac{27734375 \ln 5}{504}-\frac{6591861 \ln 3}{1400}\right)\bigg] \nonumber\\
&\quad
+\frac{G^2 p^8}{r^2} \bigg[-\frac{425}{256}+\frac{315 \nu ^5}{256}+\frac{7107 \nu ^4}{256}+\frac{2625 \nu ^3}{256} \nonumber\\
&\qquad
+\nu ^2 \left(\frac{48228101}{13440}-\frac{2125906693}{378}  \ln 2+\frac{242787134673 \ln 3}{286720}+\frac{3671798828125 \ln 5}{1548288}-\frac{96889010407 \ln 7}{221184}\right) \nonumber\\
&\qquad
+\nu  \left(\frac{82224409}{53760}-\frac{249145033}{60}  \ln 2+\frac{12166079921875 \ln 5}{6193152}+\frac{96889010407 \ln 7}{884736}-\frac{103980982797 \ln 3}{229376}\right)\bigg] \nonumber\\
&\quad
+ \frac{G^3 p^6}{r^3} \bigg[-\frac{273}{64}-\frac{6607 \nu ^4}{128}-\frac{1527 \nu ^3}{8} \nonumber\\
&\qquad
+\nu ^2 \left(-\frac{5713223}{560}+\frac{105895904239 \ln 2}{13230}+\frac{2588320706587 \ln 7}{1105920}-\frac{10808816520303 \ln 3}{2007040}-\frac{28119126171875 \ln 5}{10838016}\right)  \nonumber\\
&\qquad
+\nu  \left(\frac{2922687496621 \ln 2}{132300}-\frac{4523914911}{627200}+\frac{13469503195629 \ln 3}{5734400}-\frac{2588320706587 \ln 7}{4423680}-\frac{150369012359375 \ln 5}{14450688}\right)\bigg] \nonumber\\
&\quad
+\frac{G^4 p^4}{r^4} \bigg[-\frac{165}{32}-\frac{35 \nu ^5}{64}+\frac{2055 \nu ^4}{64}+\left(\frac{56249}{192}-\frac{41 \pi ^2}{64}\right) \nu ^3 \nonumber\\
&\qquad
+\nu ^2 \bigg(\frac{72508549}{5040}-\frac{148 \ln r}{15}-\frac{27697 \pi ^2}{6144}+\frac{296 \gamma_E }{15}-\frac{9735062548 \ln 2}{33075}+\frac{14337306321183 \ln 3}{1254400}-\frac{650540498447 \ln 7}{138240}\nonumber\\
&\quad\qquad
-\frac{2701666015625 \ln 5}{1354752}\bigg) 
+\nu  \bigg(-\frac{107 \ln r}{140}-\frac{10889 \pi ^2}{4096}+\frac{976047931}{78400}+\frac{107 \gamma_E }{70}-\frac{526259559517 \ln 2}{13230}\nonumber\\
&\quad\qquad
+\frac{33874913921875 \ln 5}{1806336}+\frac{650540498447 \ln 7}{552960}-\frac{641012819877 \ln 3}{143360}\bigg)\bigg], \\
&\hat{H}_\text{6PN(4PM)}^\text{ell,iso} = \hat{H}_\text{5PN(4PM)}^\text{ell,iso} 
+ \left[\frac{33}{2048}+\frac{429 \nu ^6}{2048}-\frac{3003 \nu ^5}{2048}+\frac{3003 \nu ^4}{1024}-\frac{1287 \nu ^3}{512}+\frac{2145 \nu ^2}{2048}-\frac{429 \nu }{2048}\right] p^{14} \nonumber\\
&\quad
+\frac{G p^{12}}{r} \bigg[\frac{273}{1024}-\nu ^6-\frac{3 \nu ^5}{2}+\frac{75 \nu ^4}{4} 
+\nu ^3 \left(-\frac{218307}{140}+\frac{10834496 \ln 2}{3}+\frac{19775583 \ln 3}{140}-\frac{138671875 \ln 5}{84}\right) \nonumber\\
&\qquad
+\nu ^2 \left(\frac{10614711}{22400}-\frac{5417248}{5} \ln 2+\frac{27734375 \ln 5}{56}-\frac{59326749 \ln 3}{1400}\right) \nonumber\\
&\qquad
+\nu  \left(-\frac{1860381}{44800}+\frac{1354312 \ln 2}{15}+\frac{19775583 \ln 3}{5600}-\frac{27734375 \ln 5}{672}\right)\bigg]  \nonumber\\
&\quad
+\frac{G^2 p^{10}}{r^2} \bigg[\frac{441}{256}+\frac{693 \nu ^6}{512}+\frac{17175 \nu ^5}{512}-\frac{2505 \nu ^4}{256} \nonumber\\
&\qquad
+\nu ^3 \left(\frac{1752882443}{134400}-\frac{50772177511 \ln 2}{3780}+\frac{15140243287719 \ln 3}{2867200}+\frac{15746212109375 \ln 5}{3096576}-\frac{1065779114477 \ln 7}{442368}\right) \nonumber\\
&\qquad
+\nu ^2 \left(\frac{930216823}{107520}-\frac{188966394467 \ln 2}{7560}+\frac{147239183828125 \ln 5}{12386304}+\frac{484445052035 \ln 7}{589824}-\frac{7125985899279 \ln 3}{2293760}\right) \nonumber\\
&\qquad
+\nu  \left(-\frac{29016839}{35840}+\frac{154094423 \ln 2}{72}+\frac{527065116993 \ln 3}{2293760}-\frac{96889010407 \ln 7}{1769472}-\frac{12533579921875 \ln 5}{12386304}\right)\bigg]  \nonumber\\
&\quad
+\frac{G^3 p^8}{r^3} \bigg[\frac{2805}{512}-\frac{19425 \nu ^5}{256}-\frac{168131 \nu ^4}{512}
+\nu ^3 \bigg(-\frac{5539742599}{120960}+\frac{38790406370519 \ln 2}{1786050}+\frac{1009279694921875 \ln 5}{877879296} \nonumber\\
&\quad\qquad
+\frac{453841966033589 \ln 7}{89579520}-\frac{244047465883413 \ln 3}{10035200}\bigg) 
+\nu ^2 \bigg(-\frac{180308862367}{2822400}+\frac{116606471572979 \ln 2}{1071630}\nonumber\\
&\quad\qquad
+\frac{3680972377512689 \ln 7}{358318080}-\frac{448065058976289 \ln 3}{40140800}-\frac{181279182489765625 \ln 5}{3511517184}\bigg) 
+\nu  \bigg(-\frac{3456473588783}{304819200}\nonumber\\
&\quad\qquad
+\frac{369057536315537 \ln 2}{9185400}+\frac{607401830370627 \ln 3}{80281600}-\frac{1267373911442149 \ln 7}{429981696}-\frac{44240036362654375 \ln 5}{2341011456}\bigg)\bigg] \nonumber\\
&\quad
+ \frac{G^4 p^6}{r^4} \bigg[\frac{2275}{256}-\frac{105 \nu ^6}{128}+\frac{1855 \nu ^5}{32}+\left(\frac{146987}{192}-\frac{41 \pi ^2}{64}\right) \nu ^4 
+\nu ^3 \bigg(-\frac{74 \ln r}{5}-\frac{25729 \pi ^2}{4096}+\frac{5104603957}{60480}+\frac{148 \gamma_E }{5}\nonumber\\
&\quad\qquad
-\frac{2348423027149 \ln 2}{51030}+\frac{8674336284777 \ln 3}{286720}+\frac{250707235071713 \ln 7}{17915904}-\frac{2232609748046875 \ln 5}{125411328}\bigg) \nonumber\\
&\qquad
+\nu ^2 \bigg(\frac{197 \ln r}{140}-\frac{197 \gamma_E }{70}-\frac{104939 \pi ^2}{16384}+\frac{2714234991803}{16934400}-\frac{126132398166437 \ln 2}{1071630}+\frac{763693932388383 \ln 3}{8028160}\nonumber\\
&\quad\qquad
+\frac{204623745011171875 \ln 5}{3511517184}-\frac{4304025048065071 \ln 7}{71663616}\bigg)
+\nu  \bigg(-\frac{5827 \ln r}{1008}-\frac{2337139 \pi ^2}{25165824}+\frac{3571766093993}{76204800}\nonumber\\
&\quad\qquad
+\frac{5827 \gamma_E }{504}-\frac{616925145960877 \ln 2}{3214890}+\frac{52541416380715625 \ln 5}{585252864}+\frac{1554400159532395 \ln 7}{107495424}\nonumber\\
&\quad\qquad
-\frac{144912376553769 \ln 3}{4014080}\bigg)\bigg].
\label{Hiso6PN}
\end{align}
\end{widetext}

This Hamiltonian can be used to check the PN expansion of a bound-orbit isotropic-gauge 4PM Hamiltonian, once the latter is computed in the future. 
Currently, it only agrees with the hyperbolic-orbit Hamiltonian of Ref.~\cite{Bern:2021dqo} at 3PN order.

In Sec.~\ref{sec:energy}, we complement the 4PM Hamiltonian $H_\text{4PM}^\text{hyp}$ with bound-orbit PN corrections $\Delta H_\text{4PM(nPN)}^\text{ell}$ to get an estimate for its effect on the circular-orbit binding energy.
We obtain those bound-orbit corrections using
\begin{equation}
\Delta H_\text{4PM(nPN)}^\text{ell} = H_\text{nPN(4PM)}^\text{ell,iso} - \left. H_\text{4PM}^\text{hyp} \right|_\text{nPN},
\end{equation}
i.e., we subtract the $n$PN expansion of $H_\text{4PM}^\text{hyp}$ from the isotropic-coordinate Hamiltonian in Eq.~\eqref{Hiso6PN}.

\section{Coefficients of the 4PM-EOB Hamiltonians}
\label{app:eobcoefs}
In this Appendix, we list the coefficients of the PM-EOB Hamiltonians, in the PS gauge of Eq.~\eqref{HeffPS} and the \PSstr gauge of Eq.~\eqref{HeffPStild}.

\subsection{Hamiltonian in the PS gauge}
When matching the scattering angle calculated from the EOB Hamiltonians to the PM-expanded scattering angle in Eq.~\eqref{chiPM}, we solve for the coefficients $q_{n\text{PM}}(\gamma)$ as functions of the effective energy.

The 2PM coefficient was derived in Ref.~\cite{Damour:2017zjx}, and it reads
\begin{equation}
q_\text{2PM}(\gamma) = \frac{3 \left(5 \gamma ^2-1\right) (\Gamma -1)}{2 \Gamma }.
\end{equation}
When working up to 3PM order, as in Ref.~\cite{Antonelli:2019ytb}, it was enough to replace $\gamma$ by the Schwarzschild Hamiltonian $\hat{H}_S$.
However, at 4PM order, we need to replace $\gamma$ by the 2PM effective energy, which we take to be the 2PM expansion of Eq.~\eqref{HeffPS}, i.e.,
\begin{equation}
\gamma \to \hat{H}_S +\frac{q_\text{2PM}(\hat{H}_S)}{2 \hat{H}_S} u^2.
\end{equation}
In the 3PM and 4PM coefficients, we simply replace $\gamma$ by $\hat{H}_S$.

The 3PM coefficient is given by Eq.~(2.17) of Ref.~\cite{Antonelli:2019ytb}, which reads
\begin{align}
q_\text{3PM}(\gamma) &= \frac{8 \left(4 \gamma ^4-12 \gamma ^2-3\right) \nu }{\sqrt{\gamma ^2-1} \Gamma ^2} \sinh ^{-1}\left(\frac{\sqrt{\gamma -1}}{\sqrt{2}}\right) \nonumber\\
&\quad
+ \frac{1}{6 \left(\gamma ^2-1\right) \Gamma ^2} \Big[
9 \left(10 \gamma ^4-7 \gamma ^2+1\right) \Gamma \nonumber\\
&\quad\qquad
-9 \left(10 \gamma ^4-7 \gamma ^2+1\right) \Gamma ^2 \nonumber\\
&\quad\qquad
+8 \gamma  \left(14 \gamma ^4+11 \gamma ^2-25\right) \nu 
\Big].
\end{align}
The 4PM coefficient we obtain reads
\begin{widetext}
\begin{align}
&\frac{\Gamma ^3 \varepsilon}{\nu}q_\text{4PM}(\gamma) = \frac{7}{8} \left(380 \gamma ^3+380 \gamma ^2+169 \gamma +169\right) \mr E\left(\frac{\gamma -1}{\gamma +1}\right)^2
+\left(300 \gamma ^2+\frac{2095 \gamma }{4}+\frac{417}{2}\right) \mr K\left(\frac{\gamma -1}{\gamma +1}\right)^2 \nonumber\\
&\quad
-\left(300 \gamma ^3+665 \gamma ^2+\frac{2929 \gamma }{4}+\frac{1183}{4}\right) \mr E\left(\frac{\gamma -1}{\gamma +1}\right) \mr K\left(\frac{\gamma -1}{\gamma +1}\right)
-\frac{8 \left(12 \gamma ^6-40 \gamma ^4+3 \gamma ^2+3\right) \Gamma  \ln \left(\sqrt{\gamma -1}+\sqrt{\gamma +1}\right)}{\sqrt{\varepsilon }} \nonumber\\
&\quad
-\frac{\gamma ^2 \left(3-2 \gamma ^2\right)^2 \left(35 \gamma ^6-65 \gamma ^4+41 \gamma ^2-11\right) \ln ^2\left(\gamma +\sqrt{\gamma -1} \sqrt{\gamma +1}\right)}{8 \varepsilon ^3}
+\frac{2 \gamma  \left(75 \gamma ^8-215 \gamma ^6-143 \gamma ^4-569 \gamma ^2+852\right) \ln \gamma}{3 \varepsilon } \nonumber\\
&\quad
+\frac{1}{4} \left(-25 \gamma ^8+125 \gamma ^6+180 \gamma ^5-49 \gamma ^4+48 \gamma ^3-9 \gamma ^2-228 \gamma -42\right) \ln ^2(\gamma +1) 
+\frac{\left(\gamma ^2-1\right) \ln \varepsilon}{12 \varepsilon } \bigg[210 \gamma ^6-552 \gamma ^5 \nonumber\\
&\qquad
+3 \gamma ^4 (70 \varepsilon  \ln 2+113) +24 \gamma ^3 (15 \varepsilon  \ln 2-38)+\gamma ^2 (3148-900 \varepsilon  \ln 2)+24 \gamma  (19 \varepsilon  \ln 2-139)-30 \varepsilon  \ln 2+1151\bigg] \nonumber\\
&\quad
+\ln (\gamma +1)\ln \varepsilon \frac{1}{2} \left(-35 \gamma ^6-60 \gamma ^5+185 \gamma ^4-16 \gamma ^3-145 \gamma ^2+76 \gamma -5\right)
-\ln (\gamma +1) \frac{\left(\gamma ^2-1\right)}{6 \varepsilon }  \bigg[150 \gamma ^7-75 \gamma ^6 \varepsilon  \ln 2\nonumber\\
&\qquad
-832 \gamma ^5+18 \gamma ^4 (5 \varepsilon  \ln 2-68)+2 \gamma ^3 (90 \varepsilon  \ln 2-739)+\gamma ^2 (1053 \varepsilon  \ln 2-272)+12 \gamma  (19 \varepsilon  \ln 2-420)+4 (39 \varepsilon  \ln 2-76)\bigg]\nonumber\\
&\quad
+\ln \left(\gamma +\sqrt{\gamma -1} \sqrt{\gamma +1}\right) \bigg[
\frac{\gamma  \ln \varepsilon}{4 \varepsilon ^{3/2}} \left(70 \gamma ^8-235 \gamma ^6+277 \gamma ^4-145 \gamma ^2+33\right)
-\frac{4 \gamma  \ln (\gamma +1)}{\varepsilon ^{3/2}} \left(30 \gamma ^6-71 \gamma ^4+35 \gamma ^2+6\right)\nonumber\\
&\qquad
-\frac{\gamma  \left(2 \gamma ^4-5 \gamma ^2+3\right)}{12 \varepsilon ^{5/2}} \left(210 \gamma ^6-720 \gamma ^5+3 \gamma ^4 (70 \varepsilon  \ln 2+113)-576 \gamma ^3+\gamma ^2 (3148-900 \varepsilon  \ln 2)-3504 \gamma -30 \varepsilon  \ln 2+1151\right)\bigg] \nonumber\\
&\quad
+\frac{2 \gamma }{\sqrt{\varepsilon }} \text{Li}_2\left(-\sqrt{\frac{\gamma -1}{\gamma +1}}\right) \left(30 \gamma ^5-60 \gamma ^4-7 \gamma ^3+82 \gamma ^2-57 \gamma +12\right)
+\left(-25 \gamma ^8+55 \gamma ^6+81 \gamma ^4-91 \gamma ^2-20\right) \text{Li}_2\left(\frac{1-\gamma }{\gamma +1}\right)\nonumber\\
&\quad
-\frac{1}{2} (\gamma +1)^2 \left(25 \gamma ^6-50 \gamma ^5+20 \gamma ^4+70 \gamma ^3-\gamma ^2-52 \gamma -12\right) \text{Li}_2\left(\frac{1-\gamma }{2}\right)\nonumber\\
&\quad
-\frac{2 \gamma  \text{Li}_2\left(\sqrt{\frac{\gamma -1}{\gamma +1}}\right)}{\sqrt{\varepsilon }} \left(30 \gamma ^5-60 \gamma ^4-7 \gamma ^3+82 \gamma ^2-57 \gamma +12\right),
\end{align}
where we recall that $\varepsilon \equiv \gamma^2 - 1$.

The bound-orbit 4PN correction term $\Delta_\text{4PN}^Q$ in Eq.~\eqref{HeffPS} is given by
\begin{align}
\Delta_\text{4PN}^Q(\hat{H}_S,r) &= \frac{(\hat{H}_S^2 - 1)^3 \nu }{r^2} \left(-\frac{1027}{12}-\frac{147432}{5}  \ln 2+\frac{1399437 \ln 3}{160}+\frac{1953125 \ln 5}{288}\right)\nonumber\\
&\quad
+ \frac{(\hat{H}_S^2 - 1) ^2 \nu }{r^3} \left(-\frac{78917}{300}-\frac{14099512}{225}  \ln2+\frac{14336271 \ln 3}{800}+\frac{4296875 \ln 5}{288}\right) \nonumber\\
&\quad
+ \frac{(\hat{H}_S^2 - 1)  \nu }{r^4} \left(\frac{296 \gamma _E}{15}-\frac{27139}{75}-\frac{9766576}{225}  \ln 2+\frac{1182681 \ln 3}{100}+\frac{390625 \ln 5}{36}\right) -\frac{148 (\hat{H}_S^2 - 1)}{15 r^4}  \nu \ln r \nonumber\\
&\quad
+\frac{1}{r^5} \bigg[\nu  \left(\frac{136 \gamma _E}{3}-\frac{34499}{1800}-\frac{29917 \pi ^2}{6144}-\frac{254936}{25} \ln 2+\frac{1061181 \ln 3}{400}+\frac{390625 \ln 5}{144}\right)+\frac{9 \nu ^3}{4}\nonumber\\
&\quad\qquad
+\left(\frac{205 \pi ^2}{64}-\frac{2387}{24}\right) \nu ^2\bigg] 
-\frac{68}{3 r^5}\nu \ln r -\frac{148 (\hat{H}_S^2 - 1)  \nu }{15 r^4} \ln (\hat{H}_S^2 - 1) .
\end{align}

\subsection{Hamiltonian in the \texorpdfstring{\PSstr}{PS*} gauge}
For the EOB Hamiltonian in the gauge in Eq.~\eqref{HeffPStild}, the 2PM coefficient is given by
\begin{equation}
a_\text{2PM}(\gamma) = \frac{3 \left(5 \gamma ^2-1\right) (\Gamma -1)}{2 \Gamma \gamma^2},
\end{equation}
and we replace $\gamma$ by the 2PM-expanded effective Hamiltonian, i.e.,
\begin{equation}
\gamma \to \hat{H}_S +\frac{ a_\text{2PM}(\hat{H}_S)}{2 \hat{H}_S}\left(1+p_r^2+l^2u^2\right) u^2.
\end{equation}
The 3PM coefficient is given by
\begin{equation}
a_\text{3PM}(\gamma) = \frac{q_\text{3PM}(\gamma)}{\gamma^2} -\frac{6 \left(5 \gamma ^2-1\right) (\Gamma -1)}{\Gamma \gamma^2},
\end{equation}
where we replace $\gamma$ by $\hat{H}_S$. 
Similarly, for the 4PM coefficient, we obtain
\begin{align}
& a_\text{4PM}(\gamma) =  \frac{q_\text{4PM}(\gamma)}{\gamma^2} + 
\frac{1}{12 \gamma^4 \Gamma ^3 \varepsilon} \bigg\lbrace
9 \left(195 \gamma ^6-209 \gamma ^4+49 \gamma ^2-3\right) \Gamma ^3-18 \left(135 \gamma ^6-157 \gamma ^4+41 \gamma ^2-3\right) \Gamma ^2 \nonumber\\
&\quad
+\gamma ^2 \nu  \ln 2 \Big[75 \gamma ^8 \ln 2-300 \gamma ^7+45 \gamma ^6 \ln 2+4 \gamma ^5 (416+45 \ln 2)-3 \gamma ^4 (691 \ln 2-816)+4 \gamma ^3 (739+12 \ln 2) \nonumber\\
&\qquad
+\gamma ^2 (544+1767 \ln 2)-12 \gamma  (19 \ln 2-840)+608+186 \ln 2\Big] 
-\Gamma  \Big[896 \gamma ^7 \nu -3 \gamma ^6 \left(256 \sqrt{\varepsilon } \nu  \ln 2+225\right)+704 \gamma ^5 \nu \nonumber\\
&\qquad
+9 \gamma ^4 \left(256 \sqrt{\varepsilon } \nu  \ln 2+105\right) -1600 \gamma ^3 \nu +9 \gamma ^2 \left(64 \sqrt{\varepsilon } \nu  \ln 2-33\right)+27\Big]
+4 \gamma ^2 \nu  \ln \left(\frac{\gamma +1}{2}\right) \Big[152-75 \gamma ^7+105 \gamma ^6 \ln 2\nonumber\\
&\qquad
+4 \gamma ^5 (104+45 \ln 2)+\gamma ^4 (612-555 \ln 2)+\gamma ^3 (739+48 \ln 2)+\gamma ^2 (136+435 \ln 2)+\gamma  (2520-228 \ln 2)+15 \ln 2\Big]\nonumber\\
&\quad
+3 \gamma ^2 \left(25 \gamma ^6-100 \gamma ^4-180 \gamma ^3-51 \gamma ^2-228 \gamma -42\right) \varepsilon  \nu  \ln ^2(\gamma +1)
-3 \gamma ^2\nu  \ln ^2\left(\frac{\gamma +1}{2}\right) \Big[25 \gamma ^8-125 \gamma ^6-180 \gamma ^5+49 \gamma ^4\nonumber\\
&\qquad
-48 \gamma ^3+9 \gamma ^2+228 \gamma +42\Big]
+384 \gamma ^2 \left(-4 \gamma ^4+12 \gamma ^2+3\right) \Gamma  \sqrt{\varepsilon } \nu  \ln \left(\sqrt{\gamma -1}+\sqrt{\gamma +1}\right)\nonumber\\
&\quad
-2 \gamma ^2 \nu  \ln (\gamma +1) \Big[75 \gamma ^8 \ln 2-150 \gamma ^7 -165 \gamma ^6 \ln 2+\gamma ^5 (832-180 \ln 2)-9 \gamma ^4 (117 \ln 2-2 (68+5 \ln 2))\nonumber\\
&\qquad
 +\gamma ^3 (1478-48 \ln 2)+\gamma ^2 (272+897 \ln 2)+12 \gamma  (420+19 \ln 2)+4 (76+39 \ln 2)\Big]\bigg\rbrace.
\end{align}

The bound-orbit 4PN correction term $\Delta_\text{4PN}^A$ in Eq.~\eqref{HeffPStild} is given by
\begin{align}
\Delta_\text{4PN}^A(\hat{H}_S,r) &= \Delta_\text{4PN}^Q(\hat{H}_S,r) + u^5 \left[\left(\frac{640}{3}-\frac{41 \pi ^2}{8}\right) \nu -14 \nu ^2\right].
\end{align}
\end{widetext}

\bibliography{refs}

\end{document}